\documentclass[subscriptcorrection,upint,varvw,barcolor=white,mathalfa=cal=euler,balance,hyphenate,french,pdf-a, nolists,nofoot,nocopyright]{Format}

\usepackage{setspace}

\JourName{}
\AtBeginDocument{%
  \fancyfoot[LO]{}\fancyfoot[RO]{}\fancyfoot[LE]{}\fancyfoot[RE]{}\fancyfoot[CO]{}%
  \fancypagestyle{title}{%
    \fancyhf{}%
    \fancyfoot[CE,CO]{\thepage}%
  }%
}%

\usepackage{float}
\usepackage{multirow}
\usepackage{makecell}
\usepackage{rotating}
\usepackage{array}
\usepackage{xcolor}

\begin{document}
\doublespacing

\SetAuthorBlock{Kiarash Naghavi Khanghah}{School of Mechanical, Aerospace, and Manufacturing Engineering,\\
   University of Connecticut,\\
   Storrs, CT 06269 \\
   email: kiarash.naghavi\_khanghah@uconn.edu}

\SetAuthorBlock{Hoang Anh Nguyen}{School of Mechanical, Aerospace, and Manufacturing Engineering,\\
   University of Connecticut,\\
   Storrs, CT 06269 \\
   email: mrv24001@uconn.edu}

\SetAuthorBlock{Anna C. Doris}{Department of Mechanical Engineering,\\
   Massachusetts Institute of Technology,\\
   Cambridge, MA 02139, USA \\
   email: adoris@mit.edu}

\SetAuthorBlock{Amir Mohammad Vahedi}{School of Mechanical, Aerospace, and Manufacturing Engineering,\\
   University of Connecticut,\\
   Storrs, CT 06269 \\
   email: amir.vahedi@uconn.edu}

\SetAuthorBlock{Daniele Grandi}{Autodesk Research,\\
   The Landmark @ One Market, Ste. 400,\\
   San Francisco, CA 94105, USA \\
   email: daniele.grandi@autodesk.com}

\SetAuthorBlock{Faez Ahmed}{Department of Mechanical Engineering,\\
   Massachusetts Institute of Technology,\\
   Cambridge, MA 02139, USA \\
   email: faez@mit.edu}

\SetAuthorBlock{Hongyi Xu\CorrespondingAuthor}{School of Mechanical, Aerospace, and Manufacturing Engineering,\\
   University of Connecticut,\\
   Storrs, CT 06269 \\
   email: hongyi.3.xu@uconn.edu}

\title{MCERF: Advancing Multimodal LLM Evaluation of Engineering Documentation with Enhanced Retrieval}

\keywords{Multimodal Retrieval, Retrieval Augmented Generation, ColPali, Vision Language Models, Large Language Models, Engineering Documentation, DesignQA}

\begin{abstract}
Engineering rulebooks and technical standards contain multimodal information like dense text, tables, and illustrations that are challenging for retrieval augmented generation (RAG) systems. Building upon the DesignQA framework~\cite{doris2025designqa}, which relied on full-text ingestion and text-based retrieval, this work establishes a Multimodal ColPali Enhanced Retrieval and Reasoning Framework (MCERF), a system that couples a multimodal retriever with large language model reasoning for accurate and efficient question answering from engineering documents. The system employs ColPali, which retrieves both textual and visual information, and multiple retrieval and reasoning strategies: (i) Hybrid Lookup mode for explicit rule mentions, (ii) Vision to Text fusion for figure and table guided queries, (iii) High Reasoning LLM mode for complex multi modal questions, and (iv) SelfConsistency decision to stabilize responses. The modular framework design provides a reusable template for future multimodal systems regardless of the underlying model architecture. Furthermore, this work establishes and compares two routing approaches: a single case routing approach and an agent-based system, both of which dynamically allocate queries to optimal pipelines. Evaluation on the DesignQA benchmark illustrates that this system improves average accuracy across all tasks with a relative gain of +32.6\% from baseline RAG best results, which is a significant improvement in multimodal and reasoning-intensive tasks without complete rulebook ingestion. This shows how vision language retrieval, modular reasoning, and adaptive routing enable scalable document comprehension in engineering use cases.

MCERF is publicly available at: \href{https://github.com/kiarash99Naghavi/MCERF}{https://github.com/kiarash99Naghavi/MCERF}

\end{abstract}

\date{Version \versionno, \today}

\maketitle

\section{Introduction}
Many engineering design documents, like rulebooks, standards, and technical specifications, are multimodal. Because they integrate text, math, tables, and illustrations, they can be quite complex. Understanding and reasoning over such heterogeneous information remains a major challenge for automated systems~\cite{van2023document, pramanick2024spiqa}. Large language models (LLMs), while good at reasoning, often struggle when visual cues are essential for generating accurate answers~\cite{faysse2024colpali, yin2023vlmreview, naghavi2025multimodal}. In engineering usage, diagrams, charts, and visual layouts provide critical context influencing the meaning of technical specifications. Failing to properly incorporate such visual context with the textual data would limit the LLM's ability to assist with tasks such as rule interpretation, compliance checking, or design requirement verification~\cite{liu2024agentbench, wang2024docllm}.

The DesignQA benchmark~\cite{doris2025designqa}, derived from the Formula SAE competition\footnote{\url{https://www.fsaeonline.com}}, was introduced to evaluate how multimodal LLMs perform on question-answering tasks grounded in engineering documentation. It provides a large-scale testbed where models are required to interpret textual and visual information jointly to answer engineering-related questions. However, the original DesignQA framework relied on complete document ingestion and used relatively simple retrieval approaches~\cite{lewis2020rag}, limiting its scalability and precision when deployed in real-world engineering workflows.

Building upon DesignQA, this work establishes a multimodal framework by introducing the \textbf{Multimodal ColPali Enhanced Retrieval and Reasoning Framework (MCERF)}. It is a modular system that integrates multimodal retrieval, adaptive reasoning, and dynamic query routing. Unlike multimodal full document ingestion methods, MCERF operates on retrieved multimodal content, significantly reducing computational cost while improving interpretability. The framework leverages the ColPali retriever~\cite{faysse2024colpali}, which represents both textual and visual structures, and couples it with multiple reasoning strategies:
(i) a \textit{Hybrid Lookup Mode} for explicit rule mentions,
(ii) a \textit{Vision-to-Text Fusion} for figure- and table-guided reasoning,
(iii) a \textit{High-Reasoning LLM Mode} for complex questions requiring more reasoning~\cite{wei2022cot}, and
(iv) an \textit{SelfConsistency Decision Layer} to improve stability.

We further propose adaptive routing strategies, including a single case and an agent-based routing mechanism that dynamically allocates queries to the most suitable retrieval-reasoning pipeline. These routing methods allow MCERF to balance accuracy and efficiency, adapting to the complexity of each question and task.

Comprehensive evaluations on the DesignQA benchmark demonstrate that MCERF substantially outperforms baseline retrieval augmented generation (RAG) systems. Our framework achieves the best performance across all six benchmark problem types, yielding a 32.6\% overall relative improvement over the previous best-performing baseline RAG. This indicates that integrating multimodal retrieval with adaptive reasoning pipelines enables more scalable and accurate comprehension of engineering documents.

Overall, this work advances the DesignQA text-based retrieval framework toward a practical and modular framework capable of reading and reasoning over real-world multimodal engineering documentation. Beyond Formula SAE, these methods have broader implications for intelligent design assistants, compliance checking, and technical documentation analysis across engineering domains. The key contributions are designed to be model-agnostic: (i) a modular multimodal retrieval-reasoning interface that separates retriever, reasoner, and router, (ii) evidence that maintaining document layout/visual structure during retrieval is a dominant consideration in building QA accuracy, and (iii) routing and specialization patterns (Hybrid, Vision2Text, SelfConsistency) that continue to apply even as foundation models change.
The rest of the paper is structured as follows.
Section~\ref{sec:BackGround} describes the DesignQA dataset and RAG background.
Section~\ref{sec:Methodology} presents the proposed MCERF methodology, including the multimodal retriever, reasoning strategies, and adaptive routing components.
Section~\ref{sec:results} discusses the comparative results of different RAG techniques and highlights the performance of MCERF across various tasks.
Section~\ref{sec:Future Work} outlines limitations and future research directions.  Finally, Section~\ref{sec:Conclusions} concludes the paper with key findings and insights, focusing on improved visual reasoning and more efficient multimodal retrieval frameworks for engineering applications.

\section{Background} \label{sec:BackGround}

\subsection{DesignQA Benchmark}
The DesignQA benchmark~\cite{doris2025designqa} evaluates the capacity of MLLMs to comprehend lengthy engineering documentation and integrate visual and textual information when answering queries. The benchmark was based on the Formula SAE student competition, in which a student team designs and builds a race vehicle according to a set of rules. The 1449 question-answer pairs in the benchmark are derived from the 140-page Formula SAE rulebook and real design data from the MIT Motorsports team, in an effort to capture real-world questions that an engineer might pose to an MLLM. The question-answer pairs in the benchmark are organized into six categories (Retrieval, Compilation, Definition, Presence, Dimension, and Functional Performance), each corresponding to a common task an engineer might perform when designing according to technical documentation. Each question category has an associated automatic evaluation metric. For each question, the evaluated model is provided with relevant context from the Formula SAE rulebook.

\textbf{Retrieval} questions (scored using F1 bag-of-words) require the model to reproduce, verbatim, the text of the rule corresponding to a given rule number.
\textbf{Compilation} questions (scored using F1 over the rule numbers) ask the model to assemble a list of all rule numbers related to a particular vehicle term (e.g., "suspension").
\textbf{Definition} questions (scored using bag-of-characters F1) test the model's ability to identify the name of a vehicle component highlighted in pink within a multi-view CAD rendering.
\textbf{Presence} questions (scored on yes/no accuracy) assess the model's ability to identify whether a specified component (e.g., main hoop) appears in a close-up CAD image.
\textbf{Dimension} questions (scored on yes/no accuracy) ask the model to evaluate whether an engineering drawing complies with a particular rule.
Finally, \textbf{Functional Performance} questions (scored on yes/no accuracy) present the model with an image related to design performance (e.g., FEA results) and ask whether it complies with a specific rule.
Examples of all six question types are provided in the Appendix~\ref{DE}.

Doris et al.~\cite{doris2025designqa} evaluated state-of-the-art MLLMs (at the time of writing) on the DesignQA benchmark, including \texttt{gpt-4o}~\cite{openai2024gpt4o} (GPT-4o), OpenAI's \texttt{gpt-4-1106-vision-preview}~\cite{openai2024gpt4} (GPT-4), Google AI's \texttt{models/gemini-1.0-pro-vision}~\cite{google2024gemini} (Gemini-1.0), and Anthropic's \texttt{claude-3-opus-20240229}~\cite{anthropic2024claude3opus} (Claude-Opus) and, \texttt{llava-1.5-13b}~\cite{liu2023llava15} (LLaVA-1.5). Models were evaluated using two context conditions: \textit{All-Rules}, where the entire 140-page FSAE rulebook (approximately 70,091 tokens) was provided to a model via its context window (if the model's maximum context limit was large enough), and \textit{RAG}, where only the top-15 (or top-12 for Compliance questions) most relevant document chunks were retrieved using a simple LlamaIndex implementation with OpenAI's \textit{text-embedding-3-large}. The simple RAG indexed the rulebook into 250-token chunks with 50-token overlap, and cosine similarity between question embeddings and chunk embeddings determined retrieval.

Overall, the GPT-4o-AllRules model (GPT-4o given the entire rule document in its context window for each question) was the best performing MLLM of those tested. However, providing an 140-page document to a model via its context window can prove costly, in some cases as much as 25 times more expensive than providing portions of the document via RAG~\cite{doris2025designqa}. While significantly less expensive, the models that received the rulebook context via RAG performed significantly worse on the benchmark when compared with their corresponding \textbf{AllRules} variants, indicating that the simple RAG framework struggled to provide relevant portions of the rulebook to the model. This problem of ineffective RAG serves as motivation for our work, in which we develop MCERF framework that effectively furnishes MLLMs with relevant sections of the engineering documentation.

\subsection{Retrieval Augmented Generation (RAG)}

The core idea of all RAG methods is to retrieve relevant information from domain-specific knowledge bases and use it during generation, preventing the LLM's potentially outdated or incomplete internal knowledge from being the primary source~\cite{mahdi2025ask, xu2024llm}. This approach effectively mitigates hallucinations by grounding model responses in relevant external context~\cite{shuster2021retrieval, khanghah2026zero,naghavi2025large}. The quality of retrieved content directly impacts answer accuracy, particularly when visual information is incorporated alongside text. Multimodal RAG systems leverage this principle by retrieving and integrating both textual and visual data, enabling more comprehensive and accurate response generation~\cite{joshi2024robust}.

According to Abootorabi et al.~\cite{mahdi2025ask}, multimodal retrieval strategies can be categorized into several key approaches. The first category, \textit{Efficient Search and Similarity Retrieval}, establishes a unified embedding space for retrieval. Within this method, CLIP (Contrastive Language-Image Pre-training)~\cite{radford2021learning}-based methods have emerged as the predominant approach for aligning visual and textual modalities through contrastive learning~\cite{chen2024contrastive, lee2022uniclip}. Other methods such as BLIP~\cite{li2022blip, li2023blip}, which resulted in better text and image features alignment and contrastive retrieval frameworks such as MARVEL~\cite{zhou2024marvel} and Uni-IR~\cite{wei2024uniir}, which further refine cross-modal alignment through advanced negative mining, extend this category.

The second major category, \textit{Modality-Based Retrieval}, leverages techniques to enhance retrieval efficiency by exploiting the distinctive characteristics of each modality. This category encompasses several classes. Some of them include: \textit{Text-centric retrieval}, which includes methods such as BM25~\cite{robertson2009probabilistic} (utilized in Section~\ref{GPT-5-MCERF-Hybrid-section}), BGE-M3~\cite{chen2024m3}, and ColBERT~\cite{khattab2020colbert}, the latter of which implements token-level interaction mechanisms for semantic matching; \textit{Vision-centric retrieval} utilizes systems such as ImgRet~\cite{shohan2024xl} and EchoSight~\cite{yan2024echosight} to retrieve semantically similar images based on visual query representations. Of particular relevance to the present work is the category of \textit{Document Retrieval and Layout Understanding}, which processes complete documents by integrating textual, visual, and spatial layout information. ColPali~\cite{faysse2024colpali}, which bridges both the \textit{Efficient Search} and \textit{Modality-Centric} categories, employs a patch-based approach using vision-language models to encode document pages (detailed in Section~\ref{Colpali}). Subsequent developments, including ColQwen2~\cite{wang2024qwen2,faysse2024colpali} and M3DocVQA~\cite{cho2024m3docrag}, build upon this foundation by extending the patch-based technique.

\section{Methodology} \label{sec:Methodology}

Engineering design according to technical requirements is an iterative process that includes rule discovery, design synthesis, and compliance verification~\cite{kossiakoff2011systems, zhang2026desagent}. In each step, engineers must find relevant requirements, understand technical specifications, and ensure that their designs conform to requirements. The proposed MCERF addresses the first and third steps, i.e., rule discovery and compliance verification, by automating relevant document retrieval and preliminary compliance verification. This does not imply that it replaces the judgment of an experienced engineer; instead, it saves them time in finding relevant documents and preliminary verification. This leaves more time for design synthesis. However, it is important for engineers to understand that there may be failures such as retrieval errors, visual misinterpretations, and numerical reasoning errors, as discussed in Section~\ref{sec:results}. Therefore, it is crucial that MCERF is used as an aid rather than an authority~\cite{massoudi2026agentic}.

\subsection{Framework Overview} 
\begin{figure}
\centering\includegraphics[width=1\linewidth]{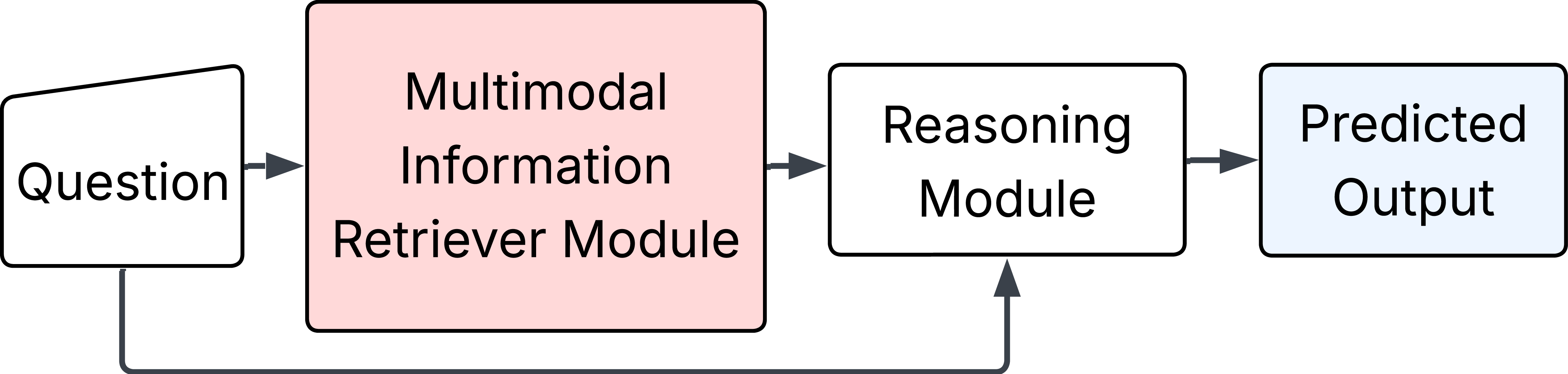}
\caption{The utilized framework containing multimodal retriever and the reasoning module\label{fig:1}}
\end{figure}

The Multimodal ColPali-Enhanced Retrieval and Reasoning Framework was proposed to facilitate question answering using engineering rulebooks and technical documents. This framework integrates a multimodal retrieval module with an LLM-based reasoning module to interpret textual and visual data. Figure~\ref{fig:1} illustrates the framework architecture (\texttt{GPT-5-MCERF-Main}), which is further detailed in subsequent sections. Furthermore, an open-source version of this framework is also available and described in Appendix~\ref{sec:Open-Source}.
\subsubsection {Multimodal Information Retriever Module} \label{Colpali}
\begin{figure}[h]
\centering\includegraphics[width=1\linewidth]{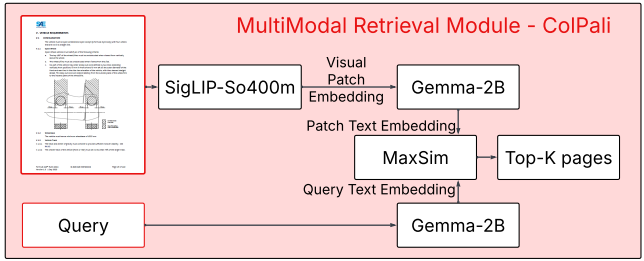}
\caption{Multimodal information retriever module framework\label{fig:2}}
\end{figure}

Engineering documents usually contain critical multimodal information (e.g., stress-strain graphs, dimensioned drawings, etc.) that must be jointly interpreted. Text-only RAGs (like the one used in DesignQA \cite{doris2025designqa}) sometimes fail to capture this multimodal information, which limits their effectiveness. To solve this issue, we employed the ColPali framework, developed by Faysse et al. \cite{faysse2024colpali}, for document indexing and query matching. ColPali operates by treating each PDF page as a discrete visual input, breaking it into smaller patches. These patch embeddings will turn into a unified representational space that maintains both textual semantics and visual attributes. When a user submits a search query, that text gets embedded using the same underlying framework. Then, a similarity matching process (late interaction \cite{lin2023fine}) compares each term from the query against all the encoded page regions, calculating similarity scores to determine relevance. This patch-level matching enables finer relevance calculation, particularly in text-heavy documents. Unlike methods such as CLIP~\cite{radford2021learning}, which generate a single embedding for an entire image and for an entire text caption, ColPali's patch-based approach captures more detail and improves the ranking quality of retrieved multimodal information~\cite{faysse2024colpali}. Figure~\ref{fig:2} demonstrates the workflow of ColPali, which extracts patch visual embeddings via SigLIP~\cite{tschannen2025siglip}, maps them to patch text embeddings through Gemma-2B~\cite{team2024gemma} along with query text encoding, and uses MaxSim scoring~\cite{lin2023fine, khattab2020colbert} to compute query-document similarity for multimodal retrieval. In this work, we employ the introduced pre-trained models without additional fine-tuning.

\subsubsection{Reasoning Module} The reasoning module employs GPT-5-mini~\cite{openai2025gpt5} as the primary language model. The input to the reasoning module consists of the textual question from the DesignQA dataset, along with any associated visual content and the multimodal retrieved context from the retrieval module, which gives the model the ability to generate answers grounded in retrieved information rather than relying on pretrained knowledge. To showcase the contribution of MCERF's multimodal retrieval, a variant was presented that utilized GPT-4o~\cite{openai2024gpt4o} as the reasoning model, allowing direct comparison with the original DesignQA baseline that used GPT-4o with a different retrieval strategy.

\subsection{Framework Variants} It is possible to enhance the accuracy and efficiency of LLMs by altering the retrieval or reasoning system, the prompt structure, and the input information \cite{wei2022chain,vatsal2024survey}. In this study, the input query contains text-heavy information with recurring keywords, while the prompted images could also include key textual elements. Four main variants of the proposed framework are introduced to improve overall performance.

\subsubsection{Variant A: \texttt{GPT-5-MCERF-Hybrid}}~\label{GPT-5-MCERF-Hybrid-section}
\begin{figure}[h]
    \centering
    \includegraphics[width=\linewidth]{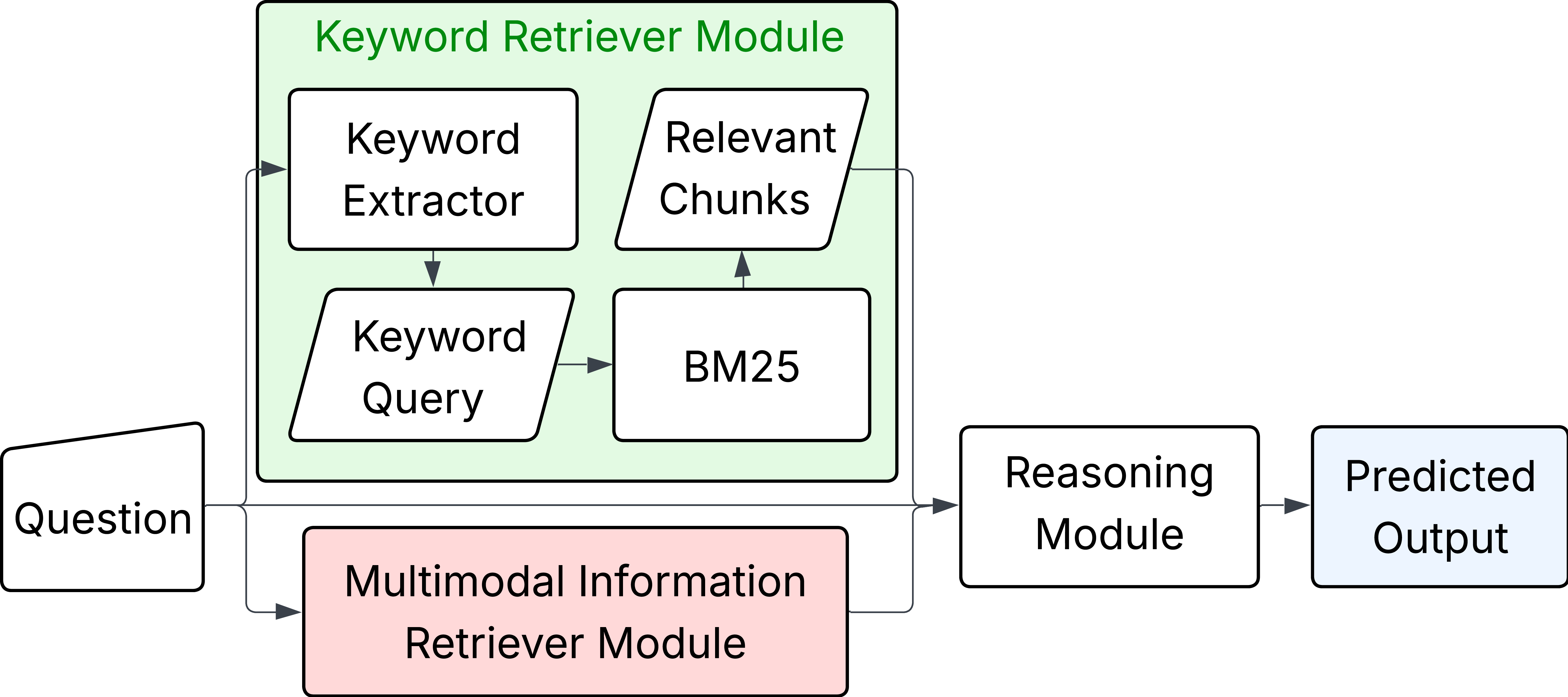}
    \caption{Hybrid Retrieval Variant combining multimodal and keyword search}
    \label{fig:3}
\end{figure}

Some DesignQA benchmark problems, such as rule extraction, depend heavily on specific terms or phrases. Since these cases often hinge on locating a particular word in the text, it is useful to include a keyword-based search alongside the semantic search currently in use \cite{chihaia2025keyword}. As illustrated in Fig.~\ref{fig:3}, the retrieval stage builds directly upon the Multimodal Information Retriever Module described in the previous section, and a Keyword Retriever Module is integrated in parallel. This Keyword Retriever Module uses an LLM (GPT-5-Nano) as a keyword extractor. Given the input question, the LLM is prompted to identify and output only the most critical technical terms, constraints, and identifiers. These extracted keywords are then used to perform a precise lexical search via the BM25 algorithm \cite{robertson2009probabilistic}, ensuring that chunks containing exact word matches are prioritized. Working in parallel, the Retriever Module captures semantic relationships across different modalities. Finally, the outputs from both the keyword-based and semantic-based retrievers are gathered and then passed to the Reasoning Module for final prediction.

\subsubsection{Variant B: \texttt{GPT-5-MCERF-SelfConsistency}} 
\begin{figure}[h]
    \centering
    \includegraphics[width=\linewidth]{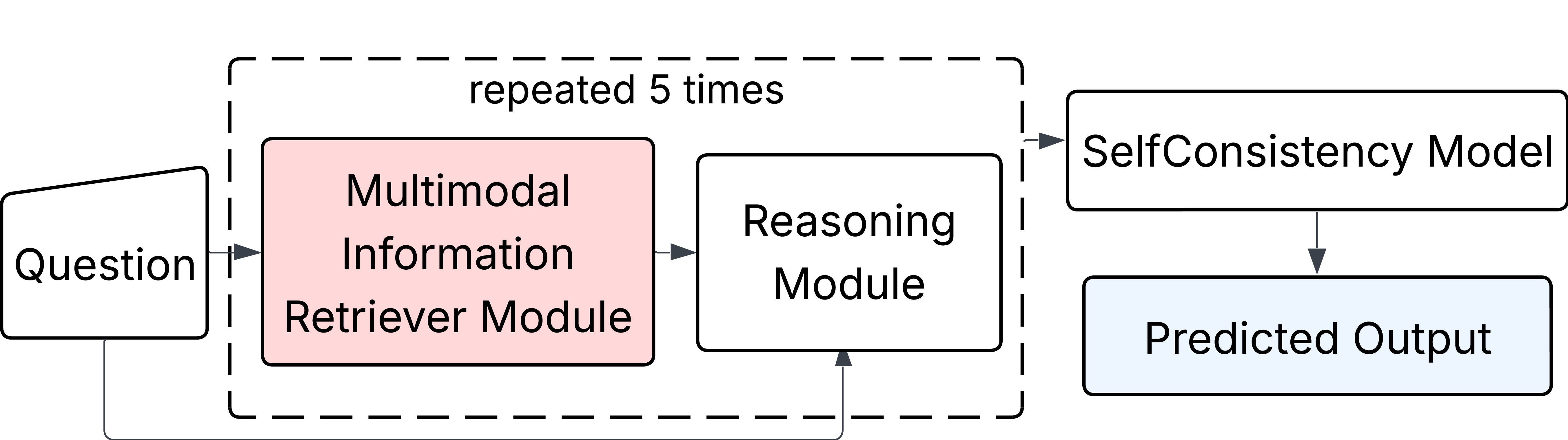}
    \caption{SelfConsistency Variant generating multiple independent retrieval--reasoning passes, where a blind adjudicator LLM consolidates results via consensus ranking to enhance robustness.}
    \label{fig:SelfConsistency}
\end{figure}

A task such as Compilation requires retrieving many rules and using them in the generation of an answer to a single question. Because there are so many rules that need to be included in the response, the reasoner module might generate slightly different answers each time it runs, even with the same input, due to the non-deterministic nature of the reasoning model used in this study~\cite{ouyang2025empirical, atil2025non}. LLMs aggregation has shown improvements in the results of generated answers and  reducing hallucination~\cite{dey2025uncertainty, yang2023one}. The architecture executes five independent retrieval--reasoning passes for every question with the default LLM in the design as shown in Fig.~\ref{fig:SelfConsistency}. In each iteration, the Multimodal Information Retriever Module retrieves relevant context, further processed by the Reasoning Module to yield a candidate answer. Therefore, this repeated sampling generates multiple response candidates that might include different aspects of the retrieved information or emphasize different reasoning paths. The candidate's answers are aggregated using a SelfConsistency Model that uses an adjudicator LLM (GPT-5-Mini), blind to the original question, seeing only the generated answers. Hence, this design inherently uses a very critical constraint that the adjudicator should not lean on its internal knowledge base in order to generate responses, but must synthesize its output solely from the presented candidates.

\subsubsection{Variant C: \texttt{GPT-5-MCERF-HighReasoning}}\label{sec:case_c}
The high-reasoning version of GPT-5-mini is used in this mode to handle tasks requiring more complex logical reasoning, such as Presence, Dimension, Functional Performance, and Definition. By extending its internal reasoning chains, the model has demonstrated improved performance on complex and multimodal problems. High reasoning model has been shown in previous research to significantly increase model accuracy in challenging tasks in need of spatial reasoning~\cite{cai2025has}. Accordingly, this variant was evaluated to quantify its potential advantages over the base model.

\subsubsection{Variant D: \texttt{GPT-5-MCERF-Vision2Text}}  
\begin{figure}[h]
    \centering
    \includegraphics[width=\linewidth]{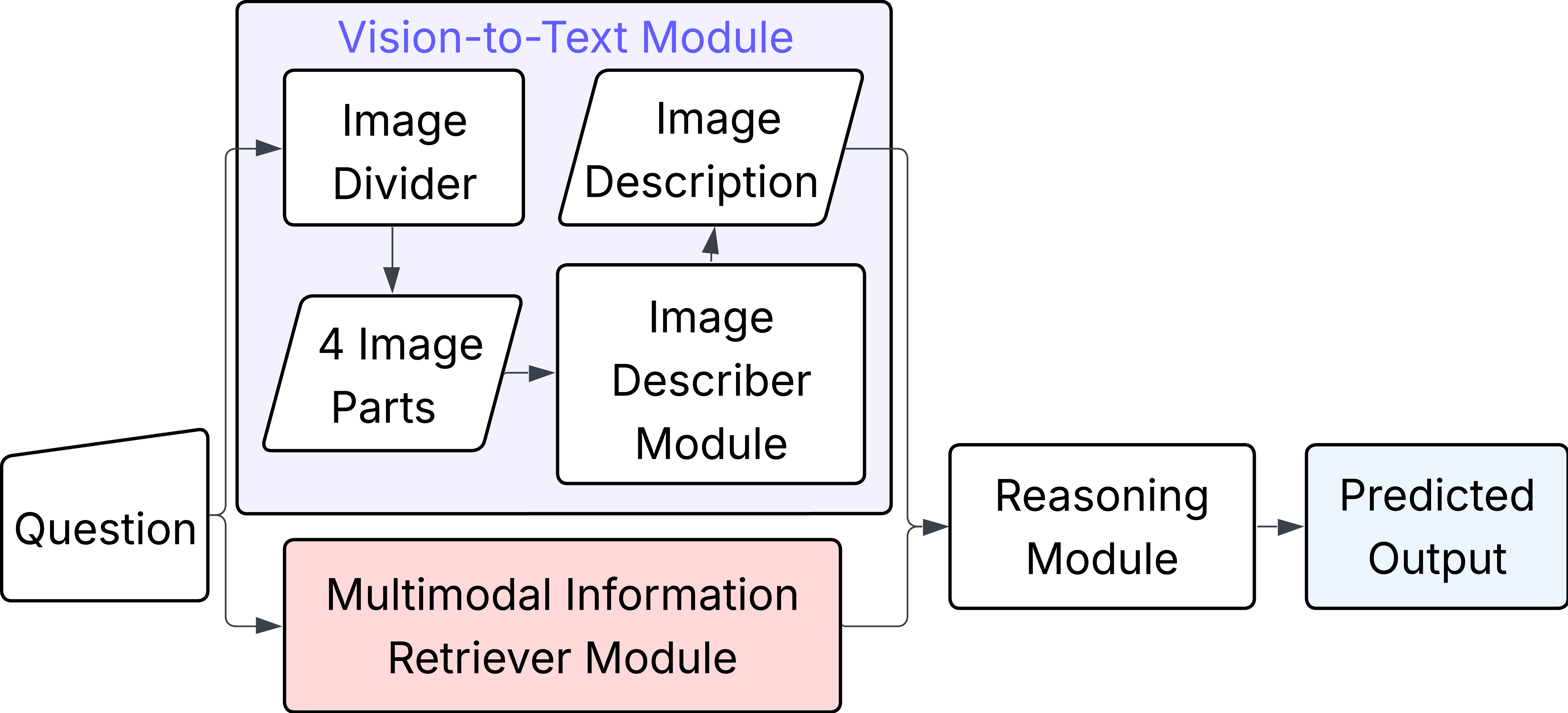}
    \caption{Architecture of the vision-to-text module. Images are segmented into overlapping quadrants, upscaled for detail preservation, and converted to textual descriptions via an image-to-text describer.}
\label{fig:V2T}
\end{figure}

A multimodal LLM's cross-image reasoning can be affected, and overall performance may decrease if it is simultaneously fed complex visual prompt inputs (query) and multimodal information sources \cite{wang2024comprehensive}. To mitigate this, a vision-to-text module (Figure~\ref{fig:V2T}) is introduced to convert visual information into textual form before reasoning.
To capture local details, each image is first split into four overlapping quadrants. Each quadrant will then be upscaled until its shortest dimension reaches at least 700 pixels in order to improve the visibility of fine-grained features~\cite{zhang2023towards}. The quadrants are then passed to an image-to-text describer (GPT-5-Mini, prompt is provided in Table~\ref{prompt:vision_to_text}), which generates detailed textual representations of visual content. These textual descriptions, combined with retrieved contextual information and question (including the original image), will be sent to the reasoning module in high-reasoning mode (as in \ref{sec:case_c}). This input transformation is intended to reduce multimodal input complexity and improve the model's ability to incorporate visual evidence into accurate final responses.

\begin{table}[t]
\caption{Vision-to-Text Prompt for image analysis}\label{prompt:vision_to_text}%
\centering{%
\footnotesize
\begin{tabular}{!{\hspace*{0.3cm}} >{\raggedright\arraybackslash} p{6.8cm} !{\hspace*{0.3cm}}}
\toprule
\textbf{System Prompt} \\ \midrule
You are a meticulous vision-language assistant. Your goal is to describe the provided plot image in such detail that someone who cannot see it could still fully understand what it shows. \par
\textit{Your description must include:} \par
\textbf{1. Overall figure:} type of chart (line, bar, scatter, etc.), title (if readable), and general layout (single panel, multiple subplots, presence of colorbars). \par
\textbf{2. Axes:} Labels (exact text if legible, else say ``unreadable''); Units (e.g., ``mm'', ``seconds'', ``\textdegree C'') or state ``not specified''; Axis ranges and tick values; Whether axes are linear, logarithmic, categorical, etc. \par
\textbf{3. Data series:} How many series are present; Their styles (color, marker, line type); Any labels in the legend; Description of each series' trend (e.g., rising, flat, peaks, correlations). \par
\textbf{4. Annotations and extras:} Text labels, arrows, highlighted regions, error bars, shading; Gridlines, secondary axes, insets, or unusual features. \par
\textbf{5. Trends \& insights:} Main relationships between x and y; Notable thresholds, turning points, or crossings between series; Comparative analysis of series. \par
\textbf{6. Uncertainties \& missing info:} If any text, axis labels, ticks, or legend entries are unreadable, state this; Mention what information does not make sense based only on the image. Avoid speculation beyond the image. \par
\textbf{7. Conclusions:} All key takeaways from the plot. \par
\textit{Output format:} JSON: structured JSON with all above categories. Report: A detailed narrative (400--700 words) accessible to someone who cannot see the figure. \\ \midrule
\textbf{User Prompt} \\ \midrule
Please analyze this plot image with the above instructions. I have attached the original figure and four zoomed quadrant crops (top-left, top-right, bottom-left, bottom-right). Use all provided views. \\ \bottomrule
\end{tabular}
}%
\end{table}

\subsection{Dynamic Model Selection} \label{DynamicModel}
Since different architectural variants perform differently on different task types, we developed an automated router that assesses question-specific features and then dynamically selects the optimal processing pipeline for each question. It reduces manual work in choosing between variants and ensures that each query is handled by the most suitable system configuration. Based on the result presented in section~\ref{sec:results}, the main framework and 3 variants (high reasoning, hybrid, and Vision to text) have been chosen to be the choices for our router. In this study, we implemented two variants of the router: Single-case and Agent-based routers.

\begin{figure*}[t]
\centering{
\includegraphics[width=\textwidth]{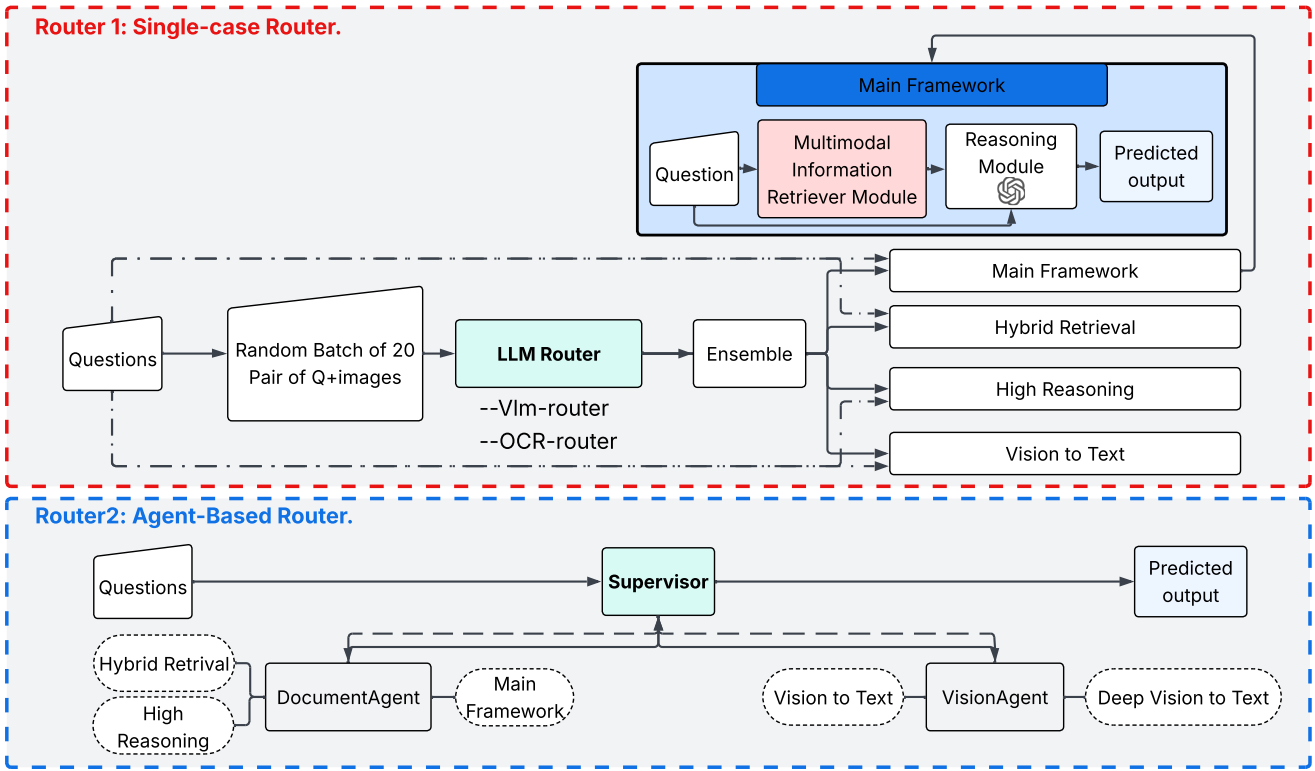}
}
\caption{Dynamic Model Selection framework illustrating both single-case and agent-based router configurations. The router automatically identifies task characteristics and activates the appropriate model pathway, removing the need for manual case selection.}
\label{fig:dynamic_router}
\end{figure*}

\subsubsection{Router1: Single-case Router} Since the questions in most subtasks share similar structures, a unified router framework was adopted. Router 1 uses LLM inference to select which pre-built variant to apply to each task type, without performing any training or tuning. For router selection, an ensemble approach was applied: up to 20 questions were randomly sampled per task (or all questions if fewer than 20). The LLM Router then queried each of these batch of each task (e.g., Definition) and determined the most suitable variant, by the final decision made by an ensemble aggregator selecting the majority-vote variant. In the end only one of the variants will be used for each task (This is task-level selection; for question-level routing, see Router 2).

\begin{table}[t]
\caption{Single-case Router Prompt}\label{prompt:routing}%
\centering{%
\footnotesize
\begin{tabular}{!{\hspace*{0.3cm}} >{\raggedright\arraybackslash} p{6.8cm} !{\hspace*{0.3cm}}}
\toprule
\textbf{System Prompt} \\ \midrule
You are a routing system for engineering QA tasks. Choose the best test script: \par
\textbf{ROUTING RULES:} \par
\textbf{1.} No image: Choose main or Hybrid  \par
\textbf{2.} Image with tables/charts/simulation results/text-heavy content → vision2text \par
\textbf{3.} Image with CAD drawings/diagrams/minimal text → high reasoning \par
\textit{Available options:} \par
\textbf{main:} For complex question requiring multiple rule finding. \par
\textbf{hybrid:} For a specific rule look up that the rule name is available. \par
\textbf{high reasoning:} For visual analysis with minimal text content such as CAD, diagrams that has minimal text \par
\textbf{vision2text:} For text-heavy technical content, tables, specifications, simulation results \par
Return JSON: \texttt{\{``test\_script'': ``option'', ``reason'': ``explanation''\}} \\ \midrule
\textbf{User Prompt (VLM-based)} \\ \midrule
Question: \texttt{\{question\}} \par
[Image attached if available] \\ \midrule
\textbf{User Prompt (OCR-based)} \\ \midrule
Question: \texttt{\{question\}} \par
Image Text: \texttt{\{image\_text\}} (or ``No image or no text in image'' if unavailable) \\ \bottomrule
\end{tabular}
}%
\end{table}

The LLM router integrates both VLM (Vision Language Model) and OCR (Optical Character Recognition) modules. The VLM processes the image--question pairs directly, while the OCR module converts images to text before feeding both the extracted text and the question into the LLM for decision-making. The full routing prompt is provided in Table~\ref{prompt:routing}.

\subsubsection{Router2: Agent-Based Router}
Router 1 operates at the task level; it identifies the best-performing variant for each task category, then applies that single variant to all questions within that task. In contrast, Router 2 operates at the question level (illustrated in Figure~\ref{fig:dynamic_router}); it analyzes each individual question and dynamically selects specialized agents based on the question's specific characteristics. Hence, different questions within the same task (e.g., Functional Performance) may be routed to different agents, each implementing the logic of different MCERF variants tailored to that question's requirements.

At the core of the architecture is the Supervisor module, which is powered by LLM. The Supervisor acts as the primary controller, receiving the initial user question and orchestrating the workflow. It is responsible for interpreting the query's intent and routing it to the appropriate agent pipeline for processing. It can assign tasks to agents repeatedly until it gets enough information to answer the original question appropriately.

The primary workflow for visual analysis involves a ``DocumentAgent'' and a ``VisionAgent'' . These agents are specialized modules designed to interpret and extract information from the original query to elements within the documents. The ``DocumentAgent'' is supported by a Hybrid Retrieval tool, which shares the same flow as Variant A, to locate keyword-based relevant information, a High Reasoning capability (Variant C) for deep analysis, and a ``Main Framework'' function related to the baseline query.

Following the creation of the framework, a specialized VisionAgent is designed to execute the question, combining image understanding process. It has control over two functions, a base ``Vision to Text'' and a ``Deep Vision to Text''. A ``Vision to Text'' module is suitable for images with tables, charts, simulation results or text-heavy contents (Variant D), while the ``Deep Vision to Text'' is prompted to adapt to hard CAD drawings, diagrams or images with minimal text that require more attempts of reasoning and the original image's visual information (Updated Variant D).

Throughout this process, either the ``DocumentAgent'' or ``VisionAgent'' provides its findings back to the Supervisor. The Supervisor's role is to synthesize the information extracted from both the textual and visual components of the document to formulate a coherent and accurate ``Predicted Output'' that directly answers the user's initial question. This agent-based, modular design allows for specialized and question level analysis.

\subsection{Evaluation Metrics}
To ensure consistency and comparability across benchmarks, the same evaluation metrics as in DesignQA~\cite{doris2025designqa} are used. These metrics eliminate the need for human judgment by enabling the automatic and impartial comparison of model predictions with ground truth answers. Every benchmark subset has a corresponding metric that corresponds to the kind and format of the questions it includes.

\textbf{F1-Based Metrics.}
Tasks such as Retrieval, Compilation, and Definition are evaluated using variants of the F1 score, a recognized metric that maintains a balance between recall and precision to represent the accuracy and completeness of the model's predictions:
\[
\text{F1} = 2 \times \frac{\text{Precision} \times \text{Recall}}{\text{Precision} + \text{Recall}}.
\]
For the Retrieval subset, F1 Bag of Words (BoW) metric has been utilized. In this method, both predicted and ground truth answers are cleaned, converted to lowercase, stripped of punctuation, and tokenized into lists of words. The overlap between these word lists is then used to calculate the F1 score, which essentially measures the amount of precise data the model extracted from the rule text.

The Compilation subset uses a closely related metric called F1 Rules, where tokens represent rule numbers rather than words. This design is suitable for questions that ask the model to extract and list specific rule identifiers instead of textual descriptions.

For the Definition subset, F1 Bag of Characters (BoC) metric has been applied. Unlike the word-level version, BoC compares character sequences, making it more tolerant of small spelling errors or variations. For instance, if the ground truth answer is "Steering tie rods," a prediction such as "Steer tie rods" is considered more accurate than "Steering column." For tasks involving component identification, this offers a more accurate assessment of similarity.

\textbf{Accuracy.}
Subsets such as Presence, Dimension, and Functional Performance are evaluated using accuracy (ACC), which measures the proportion of correct yes or no responses. These question types are treated as binary classification problems, where a prediction is correct only if it exactly matches the ground truth label.

Finally, we report macro-averaged scores across all questions within each subset. This ensures that every question contributes equally to the final score, regardless of subset size, providing a fair and balanced measure of overall model performance.

Additional metrics (BLEU, ROUGE, and Similarity) are reported for the explanation section of Rule Compliance questions; see Appendix~\ref{EM}.

\section{Results} \label{sec:results}

\begin{table*}[t]
\caption{Detailed comparison of various MLLM models' scores on DesignQA benchmark}
\label{tab:mllm_results}
\centering
\footnotesize
\setlength{\tabcolsep}{3.5pt}
\begin{tabular}{llcccccc}
\toprule
\makecell[c]{\textbf{Category}} & \makecell[c]{\textbf{Model}} &
\makecell[c]{Retrieval\\(F1 BoW ↑)} &
\makecell[c]{Compilation\\(F1 rules ↑)} &
\makecell[c]{Definition\\(F1 BoC ↑)} &
\makecell[c]{Presence\\(ACC ↑)} &
\makecell[c]{Dimension\\(ACC ↑)} &
\makecell[c]{Functional Perf.\\(ACC ↑)} \\
\midrule
\multirow{2}{*}{\makecell[l]{Baseline}}
& Baseline Naive & 0.08 & 0.14 & 0.36 & 0.50 & 0.50 & 0.50 \\
& GPT-5-NoContext & 0.05 & 0.00 & 0.60 & 0.74 & 0.30 & 0.50 \\
\midrule
\multirow{2}{*}{\makecell[l]{AllRules Models\\DesignQA~\cite{doris2025designqa}}}
& GPT-4o-AllRules & 0.88 & 0.42 & 0.54 & 0.73 & \textbf{0.83} & 0.94 \\
& GPT-4-AllRules & 0.75 & 0.30 & 0.47 & 0.63 & 0.53 & 0.56 \\
\midrule
\multirow{6}{*}{\makecell[l]{RAG Models\\DesignQA~\cite{doris2025designqa}}}
& GPT-4o-RAG & 0.19 & 0.38 & 0.53 & 0.71 & 0.68 & 0.75 \\
& GPT-5-RAG & 0.14 & 0.47 & 0.60 & 0.77 & 0.63 & 0.75 \\
& GPT-4-RAG & 0.18 & 0.36 & 0.42 & 0.53 & 0.30 & 0.54 \\
& LLaVA-1.5-RAG & 0.11 & 0.28 & 0.39 & 0.48 & 0.41 & 0.44 \\
& Gemini-1.0-RAG & 0.00 & 0.28 & 0.49 & 0.55 & 0.53 & 0.88 \\
& Claude-Opus-RAG & 0.17 & 0.29 & 0.42 & 0.51 & 0.51 & 0.88 \\
\midrule
\multirow{6}{*}{\makecell[l]{Proposed\\Framework}}
& GPT-4o-MCERF-Main & 0.61 & 0.42 & 0.54 & 0.74 & 0.75 & 0.75 \\
& GPT-5-MCERF-Main & 0.93 & \textbf{0.56} & 0.63 & 0.84 & 0.77 & 0.75 \\
& GPT-5-MCERF-Hybrid & \textbf{0.95} & 0.55 & -- & -- & -- & -- \\
& GPT-5-MCERF-SelfConsistency & 0.71 & \textbf{0.56} & 0.56 & 0.82 & 0.75 & 0.75 \\
& GPT-5-MCERF-HighReasoning & 0.92 & 0.51 & \textbf{0.64} & \textbf{0.85} & 0.80 & 0.81 \\
& GPT-5-MCERF-Vision2Text & -- & --  & 0.63 & 0.81 & 0.82 & \textbf{0.94} \\
\bottomrule
\end{tabular}
\end{table*}

\subsection{Baseline Models Results} Baseline results are obtained from the DesignQA paper~\cite{doris2025designqa}. Naive baselines represent the lower threshold for any model, as they were generated by answering the questions in a random fashion (see \cite{doris2025designqa} for details). Doris et al. tested four different state-of-the-art models (at the time of publication): OpenAI's \texttt{gpt-4o}~\cite{openai2024gpt4o} (GPT-4o), OpenAI's \texttt{gpt-4-1106-vision-preview}~\cite{openai2024gpt4} (GPT-4), Google AI's \texttt{models/gemini-1.0-pro-vision}~\cite{google2024gemini} (Gemini-1.0), and Anthropic's \texttt{claude-3-opus-20240229}~\cite{anthropic2024claude3opus} (Claude-Opus) and, \texttt{llava-1.5-13b}~\cite{liu2023llava15} (LLaVA-1.5); also in order to make comparison with the current framework, OpenAI's \texttt{GPT-5-mini}~\cite{openai2025gpt5}(GPT-5-mini) has been added as a RAG variant. Models were tested in two ways: \textit{All-Rules} models received the entire 140-page rulebook via their context windows, while \textit{RAG} models received the top-15 (or top-12 for Compliance questions) most relevant document chunks using a simple LlamaIndex RAG framework with OpenAI's \textit{text-embedding-3-large}. These results show that the simple LlamaIndex context retrieval often failed to provide the models with the necessary context for question answering, as \textit{All-Rules} models significantly outperformed \textit{RAG} variants. This performance gap, particularly notable in the Retrieval questions (for example, \texttt{GPT-4o-AllRules}: 0.89 vs. \texttt{GPT-4o-RAG}: 0.19), highlights critical limitations in simple retrieval methods. These results provide a baseline for evaluating the accuracy of our proposed retrieval approach.

\subsection{MCERF Framework Results}

Our proposed Multimodal ColPali Enhanced Retrieval and Reasoning Framework introduces several architectural innovations over the baseline RAG approaches evaluated in DesignQA. While DesignQA's Llamaindex RAG implementation retrieves top-15 most relevant chunks using cosine similarity, MCERF employs ColPali's vision-language retrieval and specialized pipelines. We evaluate five MCERF variants using GPT-4o and GPT-5-mini as backbone models, comparing their performance against DesignQA's \textit{RAG} and \textit{AllRules} baselines. Due to API cost constraints and being consistent with DesignQA's evaluation methodology, each configuration was evaluated once on the full dataset. The result is presented in Table~\ref{tab:mllm_results}. MCERF achieves an average accuracy of 0.793 across all tasks, representing a 32.6\% relative improvement over the best baseline RAG (0.598). The framework not only closes the performance gap between simple RAG and AllRules approaches but surpasses \texttt{GPT-4o-AllRules} in most cases by using specialized variants. The best-performing variants are task-dependent: \texttt{GPT-5-MCERF-Hybrid} excels at Retrieval, \texttt{GPT-5-MCERF-Main}/\texttt{GPT-5-MCERF-SelfConsistency} at Compilation, \texttt{GPT-5-MCERF-HighReasoning} at Definition and Presence, and \texttt{GPT-5-MCERF-Vision2Text} at Dimension and Functional Performance.

\begin{figure*}[t]
\centering{
\includegraphics[width=\textwidth]{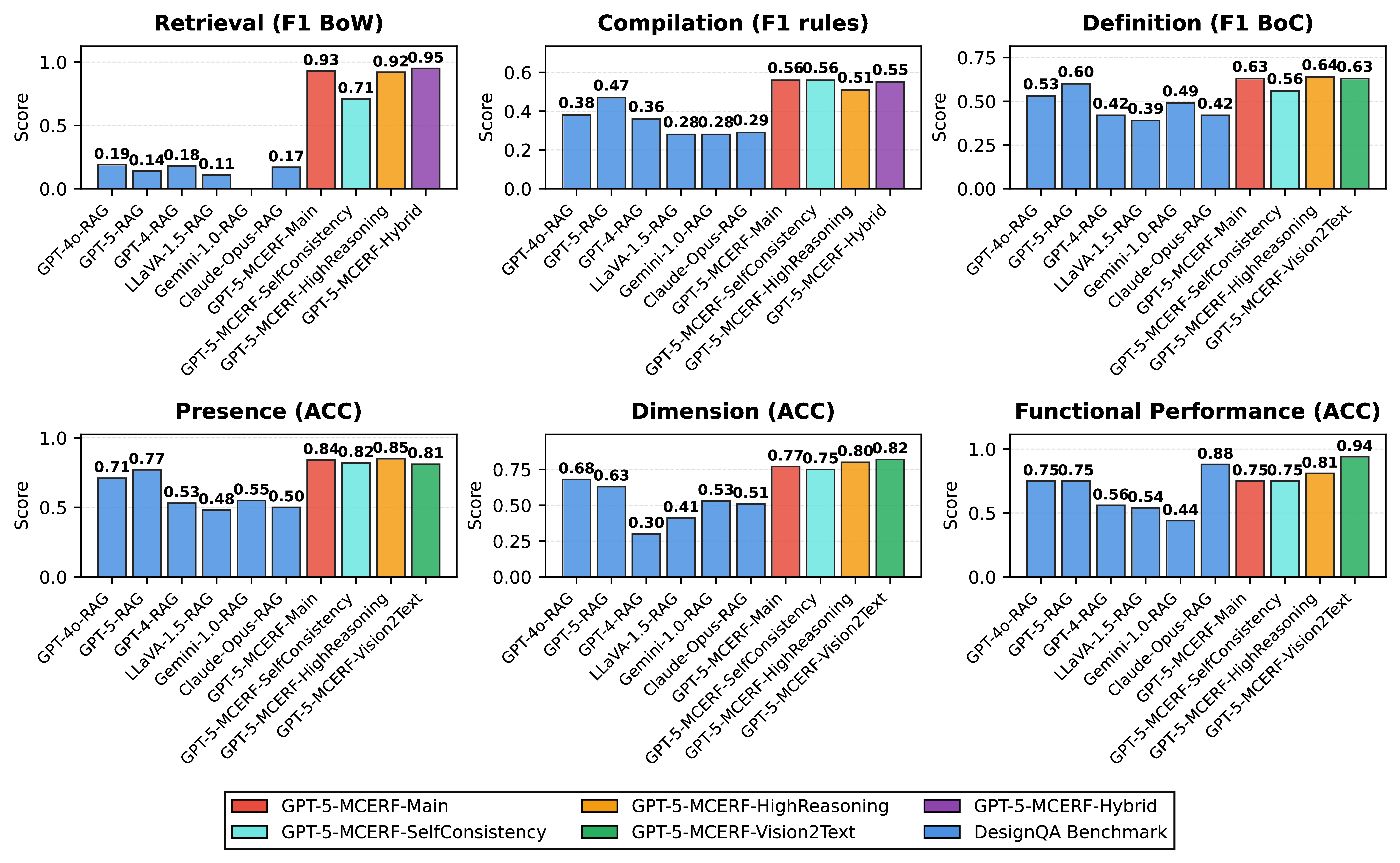}
}
\caption{Comprehensive comparison of MLLM models across six MCERF variants on DesignQA benchmark. The proposed GPT-5-MCERF framework variants (Main, SelfConsistency, HighReasoning, Vision2Text, and Hybrid) consistently outperform baseline Llamaindex RAG models across retrieval, compilation, definition, presence detection, dimension tasks, and functional performance tasks, demonstrating substantial improvements in engineering design comprehension capabilities.}
\label{fig:mllm_comparison}
\end{figure*}

\subsubsection{Retrieval (F1 BoW)}
\textbf{\texttt{GPT-5-MCERF-Hybrid}} achieves the highest score at \textbf{0.95}, followed closely by \texttt{GPT-5-MCERF-Main} (0.93) and \texttt{GPT-5-MCERF-HighReasoning} (0.92). This represents substantial improvement over \texttt{GPT-5-RAG} (0.14) / \texttt{GPT-4o-RAG} (0.19) and even surpasses the \texttt{GPT-4o-AllRules} performance (0.88) while not providing the whole context to the reasoning model.

The big improvement comes from MCERF's ColPali-based retrieval that preserves document structure. Unlike DesignQA's text extraction (LlamaIndex RAG) approach that loses formatting and hierarchy information, ColPali processes rulebook pages as images, enabling visual pattern matching over section headers and rule numbers. This improvement is highlighted by contrasting \texttt{GPT-4o-MCERF-Main} (0.61) with \texttt{GPT-4o-RAG} (0.19), and \texttt{GPT-5-MCERF-Main} (0.93) with
\texttt{GPT-5-RAG} (0.14), both sharing the same backbone model, demonstrating that ColPali's multimodal retrieval and not just model updates result in performance gains. In total, the best MCERF model obtains an \textbf{+400.0\%} gain compared to the best baseline RAG model. Also, since this task has a specific Q/A format, methods such as fine-tuning might be beneficial, which was covered in Appendix~\ref{sec:Finetune}.

\subsubsection{Compilation (F1 Rules)}
\textbf{\texttt{GPT-5-MCERF-Main}} and \textbf{\texttt{GPT-5-MCERF-SelfConsistency}} tie at the highest score of \textbf{0.56}, representing a \textbf{19.1\%} improvement over \texttt{GPT-5-RAG} (0.47). The SelfConsistency variant (\texttt{GPT-5-MCERF-SelfConsistency}) achieves 0.56 by aggregating results from 5 retrieval frameworks, compensating for individual retrieval failures that affect the results. \texttt{GPT-5-MCERF-Hybrid} (0.55) performs similarly through keyword extraction that identifies related technical terms before retrieval to perform lexical search via BM25 alongside the semantic retrieval.

\subsubsection{Definition (F1 BoC)}\label{Definition (F1 BoC)}
The highest-scoring model is \textbf{\texttt{GPT-5-MCERF-HighReasoning}} with \textbf{0.64}, a \textbf{6.7\%} improvement over the best RAG baseline, \texttt{GPT-5-RAG} (0.6). \texttt{GPT-5-MCERF-Main} and \texttt{GPT-5-MCERF-Vision2Text} are next with scores of 0.63 each. As noted by Doris et al.~\cite{doris2025designqa}, good performance on the Definition questions seems largely dependent on a model's visual reasoning capabilities, and as such, the choice of retrieval framework does not have much impact on performance (This has been proven in section~\ref{sec:Ablation}). Also, this is confirmed by the comparable accuracy of \texttt{GPT-4o-RAG} (0.53) vs \texttt{GPT-4o-MCERF-Main} (0.54), and (\texttt{GPT-5-RAG} (0.6) vs \texttt{GPT-5-MCERF-Main} (0.63)), which utilize distinct retrieval approaches (baseline RAG vs. ColPali retrieval) but the same reasoner. Hence, improvements highlighted are mostly because of upgrading the reasoner model (GPT-4o to GPT-5-mini) and leveraging specialized variants such as High Reasoning.

\subsubsection{Presence (ACC)}
\texttt{\textbf{GPT-5-MCERF-HighReasoning}} shows the accuracy of \textbf{0.85}, with GPT-5-MCERF-Main at 0.84, which is \textbf{10.4\%} better than \texttt{GPT-5-RAG} (0.77). Presence questions are a mix of visual analysis and terminology understanding. \texttt{MCERF}'s ColPali-based retrieval mitigates the context limitation identified in DesignQA: even when detailed textual descriptions are absent or limited, ColPali can leverage reference images within the rulebook to provide visual context through image-to-image matching. This provides models with the necessary terminology and visual reference information that DesignQA's text-based RAG failed to supply, and resulted in a modest increase in accuracy. Comparing models with the same reasoning module reveals GPT-4o-MCERF-Main achieves 0.74, only a 4.2\% improvement over GPT-4o-RAG (0.71), and \texttt{GPT-5-MCERF-Main} achieves 0.84, only a 9.1\% improvement over \texttt{GPT-5-RAG} (0.77). This limited improvement may be attributed to two reasons: 1) The relatively sparse visual reference content in the FSAE rulebook, suggesting that ColPali's visual retrieval capabilities are constrained by the availability of reference images in the source document. 2) Presence questions seem largely dependent on a model's visual reasoning capabilities than the retriever which has been proven in sections~\ref{sec:Ablation}.

\subsubsection{Dimension (ACC)}
\texttt{\textbf{GPT-5-MCERF-Vision2Text}} achieves the highest score at \textbf{0.82}, representing a \textbf{20.6\%} improvement over \texttt{GPT-4o-RAG} (0.68) and a 30.1\% improvement over \texttt{GPT-5-RAG} (0.63). \texttt{GPT-5-MCERF-HighReasoning} follows at 0.80. Both fall slightly short of \texttt{GPT-4o-AllRules} (0.83) but substantially exceed the RAG baselines.

Dimension questions present two challenges identified in DesignQA: (1) scale bar interpretation, where most models perform worse than with directly labeled dimensions, and (2) multi-step dimensional reasoning, where dimensions must be added or subtracted. We hypothesize that the Vision2Text pipeline outperforms other MCERF variants by addressing both challenges through a two-stage inference process. First, dimension values and their corresponding measurement locations are converted into structured text descriptions. This reduces the visual reasoning burden on the model by transforming spatial information into explicit textual relationships. Second, the reasoner performs arithmetic operations over these text-based dimension values rather than attempting to extract and compute directly from images, thereby improving accuracy on multi-step calculations. The marginal advantage of \texttt{GPT-4o-AllRules} (0.83) over \texttt{GPT-5-MCERF-Vision2Text} (0.82) likely stems from having simultaneous access to the entire 140-page rulebook, enabling it to cross-reference dimensional requirements across multiple sections without relying on retrieval. Retrieval-based methods, by contrast, are prone to errors~\cite{barnett2024seven}, such as failing to rank the most relevant pages highly enough or losing critical context. However, AllRules-based methods come at significantly higher computational cost by providing the full document to the model.

\subsubsection{Functional Performance (ACC)}
\texttt{\textbf{GPT-5-MCERF-Vision2Text}} achieves the highest score at \textbf{0.94}, representing a \textbf{6.8\%} improvement over \texttt{Claude-Opus-RAG} (0.88), which was the best RAG performer in DesignQA. Notably, \texttt{GPT-5-MCERF-Vision2Text} matches the performance of \texttt{GPT-4o-AllRules} (0.94), demonstrating that well-targeted retrieval with structured information extraction can equal full-document access. \texttt{GPT-5-MCERF-HighReasoning} also achieves strong performance at 0.81.

Functional Performance questions require integrating material properties, test results, and performance specifications with compliance rules. DesignQA finds that such questions demand ``considerable technical knowledge,'' which explains the higher performance of \texttt{Claude-Opus-RAG} over other RAG variants. Vision2Text pipeline excels here by converting heterogeneous visual formats, such as FEA stress plots and anthropometric data tables, into structured text representations that can be directly compared with rule specifications.

\subsubsection{Cross-Model Analysis}
Comparing \texttt{GPT-4o-MCERF-Main} against \texttt{GPT-5-MCERF-Main} isolates the impact of reasoning model improvements. \texttt{GPT-5-MCERF-Main} outperforms \texttt{GPT-4o-MCERF-Main}with an average improvement per task of 19.8\%, demonstrating that better language comprehension and instruction following ability enhances retrieval accuracy and multi-step aggregation tasks. In contrast, the same model upgrade within simple RAG yields only 1.9\% average improvement per task.

Why certain tasks improve under MCERF yet regress under RAG when moving to GPT-5-mini from GPT-4o is difficult to attribute to any single factor, given the black-box nature of these models~\cite{gan2026beyond}. In general, however, the average improvement across tasks suggests that the information MCERF retrieves is better suited to leveraging the capabilities of stronger reasoning models, which is reflected in the higher overall accuracy.

\subsubsection{Comparison with DesignQA Baselines}
\texttt{MCERF} significantly fills the gap in performance between RAG and AllRules approaches. Comparing in terms of GPT-4o, the macro-average gap in per-task percentage between AllRules and RAG was 71.0\%; this shrinks to only 13.2\% in MCERF with the same reasoning model and simple MCERF configuration without requiring exhaustive document ingestion. In most tasks, MCERF variants actually exceed AllRules performance, demonstrating that ColPali-based architectural improvements can compensate for reduced context when information is retrieved through vision-aware mechanisms and processed with structured reasoning. \texttt{MCERF} configuration raises the macro-average absolute score from 0.598 to 0.793, a 32.6\% relative improvement, while the macro-average of per-task percentage improvements reaches 77.3\%.

\subsubsection{Qualitative Evaluation of the Failure Cases} \label{sec:failure}

In this section, we analyze some of the most notable failure cases.

In the retrieval task, several factors contribute to drops in accuracy. A common formatting issue arises where models redundantly include the rule identifier within the prediction text (e.g., predicting ``\texttt{V.1.1} Open Wheel...'' where the ground truth is simply ``Open Wheel...''). Beyond formatting, accuracy is diminished by the listing of incorrect information, specifically in the case of the inclusion of child rules. Finally, in a limited number of cases, the models exhibit an inability to locate the required information.

In the compilation task, \texttt{GPT-4o-MCERF-Main} frequently fails to detect or retrieve all relevant rules, often missing some of the rules. In contrast, the \texttt{GPT-5-MCERF-*} models demonstrate a higher detection rate. However, these models also exhibit specific failure patterns, such as sub-rule granularity mismatches. For instance, a model may predict a set of specific sub-rules (e.g., \texttt{F.11.2.1.a}, \texttt{F.11.2.1.b}, \texttt{F.11.2.1.c}), whereas the ground truth contains only the broader parent rule (e.g., \texttt{F.11.2.1}). Consequently, this is penalized as incorrect, reducing the calculated accuracy despite the prediction being semantically relevant. Furthermore, models often exhibit parent rule redundancies or omissions. For example, a model might predict child rule (e.g., \texttt{T.7.1} and \texttt{T.7.1.1}) while missing the parent category (e.g., \texttt{T.7}). The opposite way, a model may predict a parent rule (e.g., \texttt{T.5.6}) but fail to enumerate its specific child rules (e.g., \texttt{T.5.6.2}, \texttt{T.5.6.3}, etc.), both of which negatively impact accuracy scores.

When comparing models on the definition task, as discussed in Section~\ref{Definition (F1 BoC)}, the complexity of the model's visual reasoning capabilities plays a significant role (proven in section~\ref{sec:Ablation}). Our results indicate that \texttt{GPT-4o-MCERF-Main} occasionally outputs ``I don't know'' when uncertain. In contrast, the \texttt{GPT-5-MCERF-*} models nearly always attempt an answer. While this leads to a higher rate of hallucinations, their overall performance and volume of correct outputs are superior, as detailed in Table~\ref{tab:mllm_results}.

In the dimension compliance task, the primary difficulty lies in the detection of specific components within the image that are referenced in the rule text. Accurate detection is a strict prerequisite for applying the subsequent logic; if the model fails to isolate the correct part that the provided value is referring to, it cannot verify compliance against the rule, leading to failure.

In the functional performance task, the primary challenge is the model's inability to see all numerical data and fine details, particularly when analyzing complex tables.

\subsubsection{Ablation study} \label{sec:Ablation}

To explore the extent to which the accuracy gain provided by MCERF may be due to the retrieved context rather than the model's pre-existing knowledge, an ablation study (\texttt{GPT-5-NoContext} in Table~\ref{tab:mllm_results}) has been implemented that tested the reasoner model without the retrieved information (with just the question provided in Appendix~\ref{DE}). For Retrieval, Compilation, Dimension, and Functional Performance, the model performs at or below the naive baseline without the retrieved information and the performance is much lower than MCERF models. This suggests that the model relies on the retrieved information to generate accurate answers for these questions, and we find no evidence that performance is primarily due to memorized document content. For Definition and Presence, the \texttt{GPT-5-NoContext} model performs lower than most of the \texttt{GPT-5-MCERF-*} variants but not by as large a margin. This is because questions like ``What is the name of the component highlighted in pink?''(Definition) and ``Is the front hoop visible in the close-up view?''(Presence) can be answered at least to some extent without the retrieved information from the image. Overall, these results suggest that MCERF's retrieved context  contributes to reducing hallucination and improving accuracy, though the degree of gain varies by task depending on how much each question type relies on document information.

\subsubsection{Tasks Analysis}
The performance differences across MCERF variants are not random; each task type has a distinct bottleneck that determines which pipeline is most useful. Retrieval questions are essentially a structured lookup problem rather than a semantic similarity one. The model needs to find a specific rule by its identifier, which is why adding keyword-based search in the Hybrid variant resulted in the best performance (0.95).

Compilation task requires locating all applicable rules within the document, which implies that it is a recall problem. Likely one of the reasons even the best performances obtained in this mode are modest (0.56) is that there can never be any assurance of coverage through a fixed number of retrieval pages or chunks (among other failure modes detailed in \ref{sec:failure}). However, SelfConsistency variant helps by giving each rule more chances to be captured in retrieval windows.

For Definition and Presence, as discussed in Section~\ref{sec:Ablation}, the retriever is not the bottleneck; the reasoner is. Although MCERF results in enhancement, the \texttt{GPT-5-NoContext} metrics show reasonable accuracy; therefore, to further enhance the results, these tasks actually need a stronger visual reasoning capability than additional information.

Dimension and Functional Performance are visual tasks, like Definition and Presence, but they are complex for different reasons. They ask the model to read complex figures and do arithmetic operations at the same time, which vision-language models are known to struggle with~\cite{lu2023mathvista}. The Vision-to-Text variant mitigates this by first converting the visual content into structured text as additional context, reducing the burden on the model and making the numerical information more accessible.

Stated reasons are intuition behind the routers' design in section~\ref{DynamicModel}, which route each query to the variant that best matches where its difficulty actually comes from.

\subsection{Routers Results}
\begin{figure*}[h]
\centering{
\includegraphics[width=\textwidth]{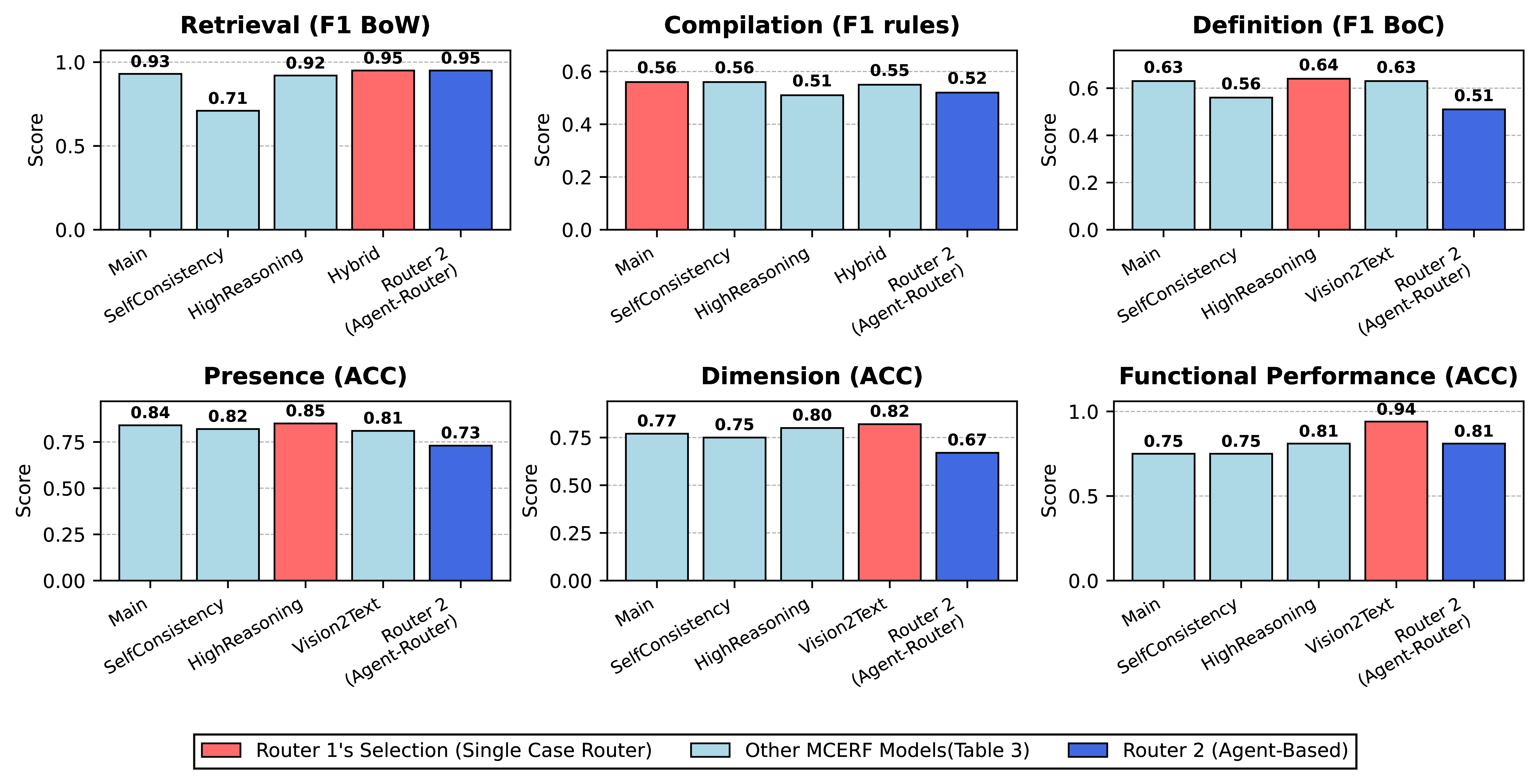}
}
\caption{Performance comparison of Single-case Router (Router 1) with the Agent-based router (Router 2) against other \texttt{GPT-5-MCERF-*} models across six evaluation tasks: Retrieval (F1 BoW), Compilation (F1 rules), Definition (F1 BoC), Presence (ACC), Dimension (ACC), and Functional Performance (ACC).}
\label{fig:router_comparison}
\end{figure*}

Evaluating a system's capability to interpret and respond accurately to queries based on complex engineering documents requires a multi-variant approach. Therefore, a router is essential to first determine the nature of a query and then direct it to the appropriate processing pipeline, whether that involves text retrieval, visual analysis, or a combination of both. As illustrated in Figure~\ref{fig:router_comparison}, Router 1 demonstrates superior performance compared to Router 2 by selecting the best variant for each task. In sections~\ref{Router 1 Results} and~\ref{Router 2 Results}, the performance of the proposed single-case router (Router 1) and agent-based framework (Router 2) will be compared against existing variants.
\subsubsection{Single-case Router (Router 1) Results} \label{Router 1 Results}

The single-case router (Router 1) demonstrated exceptional performance across multiple task categories, as illustrated in Figure~\ref{fig:router_comparison}. It strategically selects \texttt{GPT-5-MCERF-Hybrid} for Retrieval, \texttt{GPT-5-MCERF-Main} for Compilation, \texttt{GPT-5-MCERF-HighReasoning} for Definition and Presence, and \texttt{GPT-5-MCERF-Vision2Text} for Dimension and Functional Performance tasks. This routing strategy successfully identified the best-performing case for every task. The router's strong performance on text-heavy and rule-matching tasks confirms that unified routing frameworks can effectively handle the majority of engineering document queries when paired with well-designed specialized pipelines. Both OCR and LLM-based techniques give similar results, which shows the robustness of this strategy. The effectiveness of Router 1 comes from its ensemble aggregation strategy, which mitigates the risk of individual pipeline failures by sampling representative questions and selecting the most consistent processing mode through majority voting.

\subsubsection{Agent-Based Router (Router 2) Results} \label{Router 2 Results}
As shown in Figure \ref{fig:router_comparison}, the agent-based approach achieved robust capabilities and modest performance. In Retrieval (F1 BoW), Router 2 achieved a score of 0.95, indicating strong performance, similar to Router 1. Across the other metrics, such as Definition (F1 BoC), Presence (ACC), Functional Performance (ACC), Compilation (F1 rules) and Dimension (ACC), Router 2 delivered reasonable scores (0.51, 0.73, 0.81, 0.52, and 0.67, respectively). Although not always the highest-scoring model in every isolated task or narrowly optimized for a scenario, it provides balanced performance across the diverse range of tasks.

\section{Limitations and Future Work} \label{sec:Future Work}

Several promising directions emerge from this work. The most immediate one is domain adaptive retrieval. While pre-trained ColPali performs well, fine-tuning its vision encoder on engineering documents could enhance its capability to recognize domain-specific visual patterns in tasks such as dimension, presence, and functional performance. In terms of evaluation metrics other than accuracy and F1 score, Appendix~\ref{EM} demonstrates that explanations generated by both the baseline and MCERF frameworks can be evaluated by metrics such as BLEU, ROUGE, and similarity, but they achieved suboptimal values. In this case fine tuning the reasoning model on the human-generated explanation pattern might improve explanation quality (as shown in Appendix~\ref{sec:Finetune} for retrieval task), particularly for Dimension and Functional Performance questions, given enough training data.

Scaling up to massive sets of documents is both a challenge and an opportunity. Our rulebook, consisting of 140 pages, is computationally manageable, whereas real-world organizations often maintain libraries extending to thousands of pages. ColPali in MCERF demonstrates high accuracy, but its computational expenses pose scalability challenges when handling large document collections. We propose a possible solution in Appendix \ref{sec:CLIP} using CLIP-based retrieval, albeit currently at the expense of accuracy. ColPali processes rulebook pages with patch-level encoding, requiring 7.28 seconds per query for the Functional Performance task. While CLIP-based prefiltering reduces this to 6.48 seconds (11\% speedup) by processing only 30 candidate pages, it comes at the cost of a 25.1\% accuracy drop (from 0.75 to 0.562). An interesting research direction would be to develop vision retrieval methods that maintain ColPali's accuracy with increased inference speeds and lower computational costs, particularly for real-time applications.

\section{Conclusions} \label{sec:Conclusions}

In this paper, we introduce a Multimodal ColPali Enhanced Retrieval and Reasoning Framework (MCERF), a multimodal retrieval and reasoning framework achieving substantial improvement in question answering on engineering texts. We achieve an average accuracy of 0.793, a 32.6\% relative improvement over the best baseline RAG (0.598), through combining ColPali's vision-language retrieval with specialized pipelines and adaptive routing. The results demonstrate that structuring documents through patch-based vision retrieval far outperforms text-based methods. For instance, using the same reasoning model, upgrading the retrieval architecture from \texttt{GPT-4o-RAG}/\texttt{GPT-5-RAG} to \texttt{GPT-4o-MCERF-Main}/\texttt{GPT-5-MCERF-Main} increases retrieval task performance from 0.19/0.14 to 0.61/0.93, which is a 221\%/564\% improvement. Our framework settings demonstrate robust specialization: Hybrid Retrieval performs best on keyword-based tasks like retrieval (0.95 F1), Vision-to-Text performs best on dimension analysis (0.82 ACC) and functional performance (0.94 ACC), and High Reasoning performs best on complex multimodal inference (0.85 ACC on Presence and 0.64 F1 BoC on Definition). The router systems efficiently automate pipeline selection, with the multi-agent approach achieving balanced performance without manual tuning.

Several of the findings are particularly noteworthy. First, when specialized variants are considered, MCERF achieves better or comparable performance to the full-document AllRules baseline across all tasks, proving that specialized and well-designed retrieval can compensate for reduced context. Second, upgrading both the retrieval framework and the reasoning model  (from GPT-4o to GPT-5-mini) yields significant improvements on language-heavy tasks such as Retrieval and Compilation. For visual tasks, however, two distinct trends are evident: For Dimension and Functional Performance, the images are information-dense and require specialized variants (e.g., Vision-to-Text) to show enhanced performance.
For Definition and Presence, updating the retrieval framework shows modest gains on visual tasks since enhancements are mostly related to reasoner due to the nature of these question which could be answered to reasonable accuracy even without retrieved information. Third, our exploration of possibilities such as SAM segmentation (Appendix \ref{sec:SAM}), CLIP prefiltering (Appendix \ref{sec:CLIP}), and fine-tuning (Appendix \ref{sec:Finetune}), reported in appendix, actually did not lead to significant performance improvements. This suggests that modifications to the retrieval architecture and reasoning methods are more critical.

Also, engineering organizations currently face a trade-off between token-expensive full-document ingestion and accuracy-limited naive RAG. MCERF offers a third path,  approaching AllRules performance while maintaining a reasonable efficiency (Appendix~\ref{sec:Cost}). This work demonstrates that the bottleneck in engineering document understanding lies not in model capability alone, but in the preservation and strategic use of multimodal information during retrieval.

\section*{Acknowledgment}

We gratefully acknowledge the financial support from the National Science Foundation CMMI-2142290. KNK also gratefully acknowledges the Pratt \& Whitney Institute for Advanced Systems Engineering Fellowship from the University of Connecticut.

\section*{Nomenclature}

GPT-5-mini= OpenAI's GPT-5-mini model (primary reasoning LLM in MCERF)

GPT-4o= OpenAI's gpt-4o model

GPT-4= OpenAI's gpt-4-1106-vision-preview model

Claude-Opus= Anthropic's claude-3-opus-20240229 model

Gemini-1.0= Google's gemini-1.0-pro-vision model

LLaVA-1.5= Open-source llava-1.5-13b model

ColPali= Vision-language retrieval model using patch-based document encoding

MCERF= Multimodal ColPali Enhanced Retrieval and Reasoning Framework

LLM= Large Language Model

MLLM= Multimodal Large Language Model

RAG= Retrieval Augmented Generation

-RAG= Models with LlamaIndex's simple RAG framework

-AllRules= Models given entire rule document via context window

BM25= Keyword search algorithm

CLIP= Contrastive Language-Image Pre-training

FSAE= Formula SAE (student engineering competition)

\appendix

\section{Image Segmentation and Attention Refinement Study} \label{sec:SAM}
\subsection{Motivation}
To enhance the visual understanding capability of our MCERF method for the Formula SAE dataset, we explored the effect of replacing raw image inputs with segmented regions containing more informative content. The hypothesis was that large images with extensive background or empty areas might dilute the attention of the vision encoder, such as ViT, and increase computational load. Therefore, we sought to extract meaningful regions of interest (ROIs) using a segmentation model.

\subsection{Segmentation with SAM}
We utilized the Segment Anything Model (SAM)~\cite{kirillov2023segment} to segment each image into multiple subregions. The idea was to discard visually irrelevant background pixels and keep only regions with structural or textual information, such as annotated diagrams, rule schematics, and vehicle components~\cite{yao2025stepideator}. Each segmented patch was then concatenated and passed to the visual encoder (ViT) as the image input to the MCERF pipeline.

\begin{figure}[h]
    \centering
    \includegraphics[width=\linewidth]{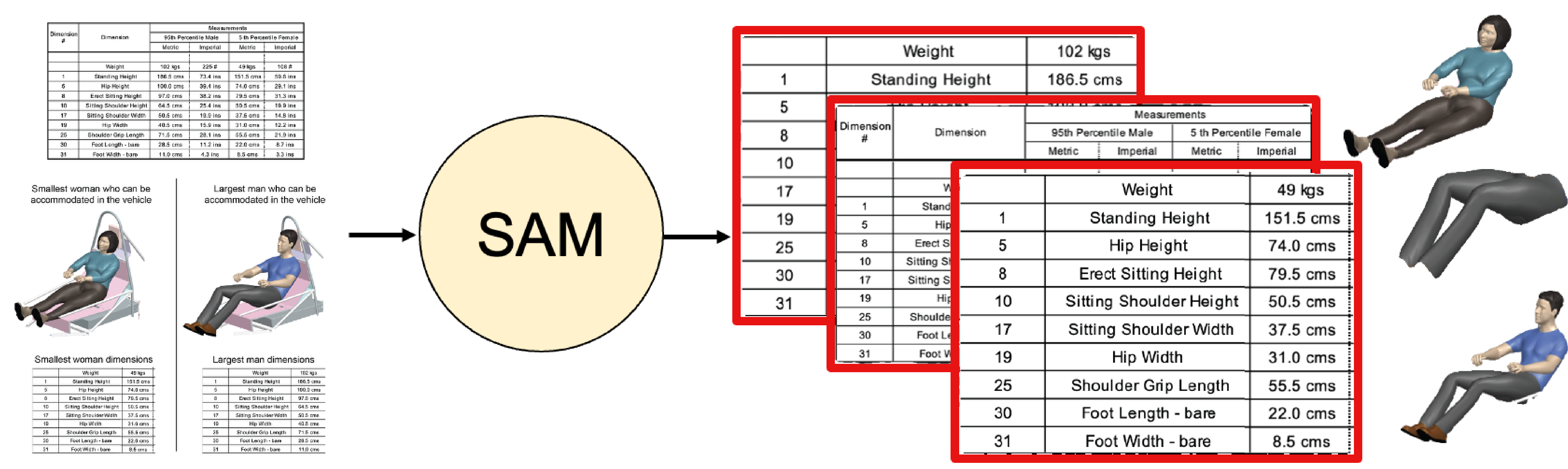}
    \caption{SAM-based image preprocessing pipeline. }
    \label{fig:sam_segmentation}
\end{figure}

\subsection{Observations and Limitations}

Despite the benefits of segmentation, there are various challenges. Segmenting images reduces the size of images and emphasizes certain areas, but critical components such as suspension elements, charts, and textual annotations were in some cases dismembered, fragmented, or omitted altogether. This resulted in several dissociated parts where the model logic failed, and resulted in poor performance.

To address this, we added the segmented image patches together with the original image to the MCERF model. The performance dropped again where overlapping or competing visual embeddings of the same image were poor in aiding attention.

Table~\ref{tab:sam_comparison} shows the evidence of this performance drop with the exception of the \textit{Definition} category where SAM provides a small improvement. In this case, it seems to aid the model in concentrating on detailed local elements such as equations or specific textual areas. In the \textit{Presence}, \textit{Dimension}, and \textit{Functional Performance} categories, the performance drop was the most severe. This was most pronounced with the Vision2Text-ColPali variant where the model once again improved, showing the implicit power of verbalization in text via visual features as an aid to contextualization, and in this case, to fragment the supporting context lost in segmentation.

\begin{table}[h]
\centering
\caption{Comparison of MCERF with and without SAM Segmentation}
\label{tab:sam_comparison}
\begin{tabular}{lc}
\toprule
\textbf{Method} & \textbf{Score} \\
\midrule
\multicolumn{2}{l}{\textit{Definition (F1 BoC)}} \\
\midrule
Best MCERF (non-SAM) & 0.64 \\
GPT5Reasoning-ColPali-SAM & 0.61 \\
GPT5Reasoning\_Vision2Text-ColPali-SAM & 0.67 \\
\midrule
\multicolumn{2}{l}{\textit{Presence (ACC)}} \\
\midrule
Best MCERF (non-SAM) & 0.85 \\
GPT5Reasoning-ColPali-SAM & 0.72 \\
GPT5Reasoning\_Vision2Text-ColPali-SAM & 0.79 \\
\midrule
\multicolumn{2}{l}{\textit{Dimension (ACC)}} \\
\midrule
Best MCERF (non-SAM) & 0.80 \\
GPT5Reasoning-ColPali-SAM & 0.43 \\
GPT5Reasoning\_Vision2Text-ColPali-SAM & 0.53 \\
\midrule
\multicolumn{2}{l}{\textit{Functional Performance (ACC)}} \\
\midrule
Best MCERF (non-SAM) & 0.94 \\
GPT5Reasoning-ColPali-SAM & 0.75 \\
GPT5Reasoning\_Vision2Text-ColPali-SAM & 0.88 \\
\bottomrule
\end{tabular}
\end{table}

Overall, using SAM-based segmentation did not improve MCERF's performance on our multimodal question-answering task. It seems that keeping the full image, with all its spatial and contextual information, helps the model reason more effectively than aggressively cutting images into smaller segmented parts.

\section{Contrastive Language-Image Pre-Training (CLIP) Filtering} \label{sec:CLIP}

ColPali processes every page by dividing it into patches, adding each patch separately, and afterwards using MaxSim to compare each query token against all of the patch embeddings as described fully in section~\ref{Colpali}. Patch-level comparison will result in fine-grained visual and text features comparisons which is more expensive than regular retrieval. Alternatively, CLIP captures the entire page into a single embedding and computes semantic similarity at the page level, which is much quicker. We test a two-stage pipeline that uses CLIP for fast initial filtering before applying ColPali's more detailed analysis. CLIP quickly finds similar pages through whole-page embeddings, then ColPali examines only those candidates with its patch-based matching.

\subsection{CLIP + ColPali Framework}
The pipeline works in two steps:
\begin{enumerate}
    \item \textbf{CLIP Prefiltering}: CLIP computes similarity scores across all 140 pages and selects the top-30 candidates. CLIP's efficient architecture makes this step fast.
    \item \textbf{ColPali Reranking}: ColPali processes only these 30 pages, computing detailed embeddings and reranking them based on fine-grained pattern matching.
\end{enumerate}
\begin{figure}[h]
    \centering
    \includegraphics[width=\linewidth]{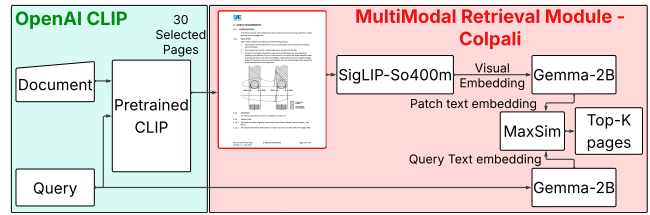}
    \caption{Two-stage retrieval architecture combining CLIP prefiltering with ColPali multimodal retrieval.}
    \label{fig:clip_colpali}
\end{figure}
\subsection{Motivation}
This cuts ColPali's workload by roughly 78.5\% (30 pages instead of 140) while keeping its superior retrieval on the most promising candidates.

\subsection{Results}
Table~\ref{tab:clip_colpali} compares the two approaches on Functional Performance questions.

\begin{table}[h]
\centering
\caption{Comparison of retrieval methods on Functional Performance}
\label{tab:clip_colpali}
\begin{tabular}{lcc}
\toprule
\textbf{Method} & \textbf{ACC} & \textbf{Time (s)} \\
\midrule
Original Method (ColPali only) & 0.75 & 7.28 \\
CLIP + ColPali & 0.562 & 6.48 \\
\bottomrule
\end{tabular}
\end{table}

CLIP prefiltering reduces latency by 11.0\% (7.28s to 6.48s) but accuracy drops 25.1\% (0.75 to 0.562). The problem is that CLIP's coarse filtering sometimes removes relevant pages from the top-30, so ColPali never sees them. The modest speedup makes sense given the small rulebook size. Processing 78.5\% fewer pages only saves about 0.8 seconds. Aerospace, automotive, and construction engineering rulebooks commonly have 500-1000+ pages. For those assignments, likewise top-30 filtering would provide 2-3x accelerations and be more precise, since CLIP would then have more clearly irrelevant pages to exclude and the candidate set would be a much smaller proportion of the entirety.

\section{Computational Cost} \label{sec:Cost}

\begin{figure}
    \centering
    \includegraphics[width=1\linewidth]{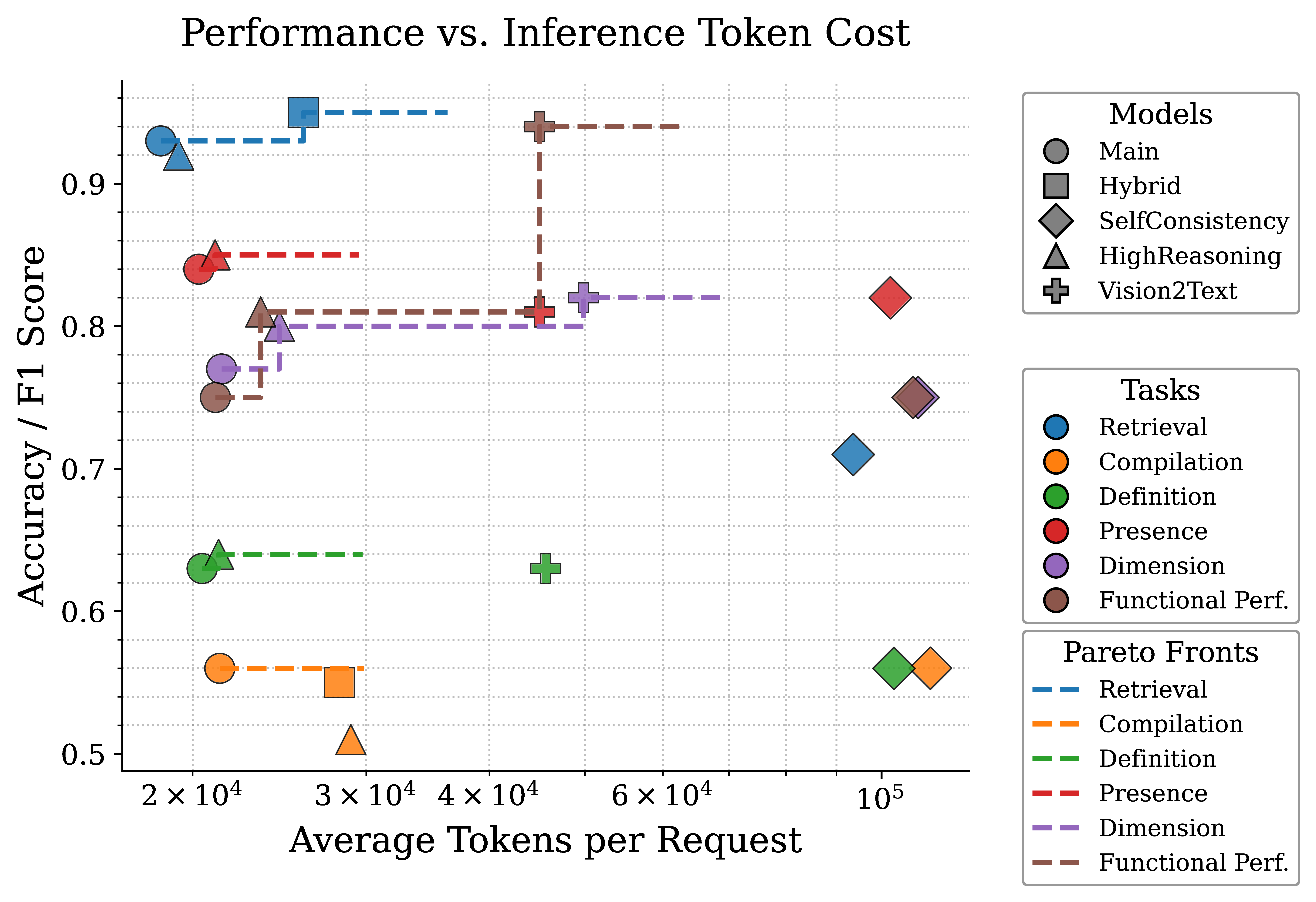}
    \caption{Performance vs. inference token cost across model variants (all using \texttt{GPT-5-MCERF-*}) and tasks. Dashed lines indicate the Pareto frontier for each task.}
    \label{fig:cost}
\end{figure}

In this section, Figure~\ref{fig:cost} presents performance against average token usage per request for each task (colors) and each model (marker shapes). To get the average token usage, 15 random questions were selected and run through each pipeline to capture token usage for both the reasoner model and helper models, such as the image describer module in Vision-to-Text. A Pareto front is also shown for each task to visualize which models achieve higher accuracy at lower cost.

\subsection{Results}

The results are broadly consistent with the findings reported throughout the paper. \texttt{GPT-5-MCERF-Main} is the most token-efficient variant, as it is also the simplest. To achieve better results, especially in cases where images are included in the questions, \texttt{GPT-5-MCERF-HighReasoning} or \texttt{GPT-5-MCERF-Vision2Text} improve accuracy at the cost of higher token usage. For a task that do not include images, such as Retrieval, the specialized \texttt{GPT-5-MCERF-Hybrid} variant achieves the best accuracy. \texttt{GPT-5-MCERF-SelfConsistency} is the most token-intensive variant, as it runs five independent retrieval-reasoning passes per question, pushing token usage to nearly $10^5$ per request. However, this added cost does not consistently translate into accuracy gains over lower-cost variants, making it a poor trade-off for most tasks.

\section{Fine-Tuning Attempt} \label{sec:Finetune}

\subsection{Motivation}
Some tasks such as Retrieval in DesignQA has a specific format: given a rule ID, the model must return the exact text of that rule without any additional explanation or context. This particular format constraint led us to reflect that fine-tuning over rule ID to rule text mappings could improve performance by training the model to produce DesignQA-consistent answers~\cite{naghavi2025reconstruction}. It is possible that fine-tuning GPT-4o over question-answer pairs directly extracted from the rulebook would remember both the reasoning form and the very formatting the benchmark expects, which results in increase in accuracy.

\subsection{Fine-Tuning Process}

The parsing pipeline identified rule IDs with specific patterns (e.g., \texttt{AA.1.1.1}, \texttt{D.13.2.2}) and assigned content exclusively to the deepest open rule to avoid overlapping text between parent and child rules. We have randomly selected approximately 2\% of our recovered rules from our ColPali pipeline and generated question-answer pairs in the format: ``What precisely does rule [ID] say? Answer with only the text of the rule and nothing more.'' We used \texttt{gpt-4o-2024-08-06} as our base fine-tuning model since it supports multimodal inference, so it is directly applicable to use with MCERF's image-based retrieval pipeline.

\subsection{Results and Analysis}
Table~\ref{tab:finetuning} compares the fine-tuned model against baseline approaches on the Retrieval task.

\begin{table}[h]
\centering
\caption{Retrieval performance comparison with fine-tuned model}
\label{tab:finetuning}
\begin{tabular}{lc}
\toprule
\textbf{Model} & \textbf{Retrieval (F1 BoW)} \\
\midrule
GPT-5-MCERF-Main & 0.93 \\
GPT-4o-MCERF-FineTuned & 0.86 \\
GPT-4o-MCERF-Main & 0.61 \\
GPT-4o-AllRules & 0.88 \\
\bottomrule
\end{tabular}
\end{table}

Fine-tuning showed improvement even with limited data. The fine-tuned model (\texttt{GPT-4o-MCERF-FineTuned}) scored 0.86 on Retrieval, a 41.0\% improvement over the original \texttt{GPT-4o-MCERF-Main} (0.61).  This shows that fine-tuning strategy successfully teaches the model to produce the exact formatting and content structure expected by the benchmark questions. With just 2\% of the rulebook rules included in the training set, the model could learn to handle rule hierarchies and precise text replication more effectively than the base ColPali retrieval pipeline alone.

However, the fine-tuned model still falls short of GPT-5 models (0.93) and \texttt{GPT-4o-AllRules} (0.88). We intentionally limited our training set to 2\% of the rulebook because we wanted to avoid providing the full rulebook as context during training, which would defeat the purpose of testing RAG-based approaches. Despite this constraint, the 41\% improvement over the base model suggests that fine-tuning remains a promising direction.

\section{Open-Source Implementation} \label{sec:Open-Source}
To demonstrate that the proposed framework works fully with open models, we used \texttt{unsloth/Llama-3.2-11B-Vision-Instruct-bnb-4bit} as the reasoning backbone in our pipeline. Table~\ref{tab:mllm_results} reports results for proprietary models; for this open-source reasoner, we observe the following scores on DesignQA: Retrieval (F1 BoW) 0.26, Compilation (F1 rules) 0.25, Definition (F1 BoC) 0.39, Presence (ACC) 0.50, Dimension (ACC) 0.60, and Functional Performance (ACC) 0.50.
\begin{table}[h]
\centering
\caption{Comparison of open-source reasoner performance}
\label{tab:opensource_vs_llava}
\footnotesize
\setlength{\tabcolsep}{4pt}
\begin{tabular}{lcccccc}
\toprule
\textbf{Model} & \textbf{Retr.} & \textbf{Comp.} & \textbf{Def.} & \textbf{Pres.} & \textbf{Dim.} & \textbf{Func.} \\
\midrule
\texttt{Llama-11B-MCERF} & 0.26 & 0.25 & 0.39 & 0.50 & 0.60 & 0.50 \\
\texttt{LLaVA-1.5-RAG} & 0.11 & 0.28 & 0.39 & 0.48 & 0.41 & 0.44 \\
\midrule
\multicolumn{7}{l}{\textit{Main Closed-Source Model}} \\
\midrule
\texttt{GPT-5-MCERF-Main} & 0.93 & 0.56 & 0.63 & 0.84 & 0.77 & 0.75 \\
\bottomrule
\end{tabular}
\end{table}

Although \texttt{Llama-11B-MCERF} model is smaller than the models used in the baselines (e.g., \texttt{LLaVA-1.5-RAG} is a 13B model) and it is also quantized, its results are comparable to (and sometimes higher than) the LlamaIndex RAG baselines, indicating that the framework remains usable under constrained compute. In this study, due to limited local computational resources, we used a relatively small (11B) 4-bit quantized model, which explains part of the accuracy drop relative to stronger proprietary backbones.

For higher performance, it is advisable to use open models with reasoning capabilities closer to the proprietary models evaluated in this paper (e.g., \texttt{moonshotai/Kimi-K2.5}). The project repository includes the open-source pipeline code, and due to the modular architecture of MCERF, swapping the reasoning model for a stronger backbone is straightforward.

\section{DesignQA QA Samples} \label{DE}
Table~\ref{tab:designqa_samples} presents examples of questions and answers for each task category.
\begin{table}[h]
\caption{Examples of DesignQA tasks: Rule Extraction, Comprehension, and Compliance.}
\label{tab:designqa_samples}
\centering
\footnotesize
\setlength{\tabcolsep}{6pt}
\resizebox{\columnwidth}{!}{%
\begin{tabular}{l p{3.0cm} p{3.0cm}}
\toprule
\makecell[l]{\textbf{Sub-Task}} &
\makecell[l]{\textbf{Question}} &
\makecell[l]{\textbf{Answer}} \\
\midrule
Retrieval & Tell me rule V.1.2 verbatim. & The vehicle must have a minimum wheelbase of 1525 mm. \\
\cmidrule{1-3}
Compilation & List all the rules relevant to suspension. & V.3.1.1, V.3.1.2, V.3.1.3, V.3.1.4, T.1.3.3, T.1.3.4, F.3.4.3, ... \\
\midrule
\makecell[l]{Functional\\Performance} & Does the design comply with F.8.7.2? Answer with an explanation and a yes/no. + \textcolor{blue}{Figure~\ref{fig:examplesQA}-A}& Explanation: The design doesn't comply... Answer: no \\
\cmidrule{1-3}
Definition & What is the name of the component highlighted in pink? + \textcolor{blue}{Figure~\ref{fig:examplesQA}-B}& chassis; frame; space frame \\
\cmidrule{1-3}
Dimension & Does the design comply with T.7.7.1a? Answer with an explanation and a yes/no. + \textcolor{blue}{Figure~\ref{fig:examplesQA}-C}& Explanation: The design complies... Answer: yes \\
\cmidrule{1-3}
Presence & Is the front hoop visible in the close up view? + \textcolor{blue}{Figure~\ref{fig:examplesQA}-D}& No \\
\bottomrule
\end{tabular}%
}
\end{table}
\begin{figure*}[h]
    \centering
    \includegraphics[width=\linewidth]{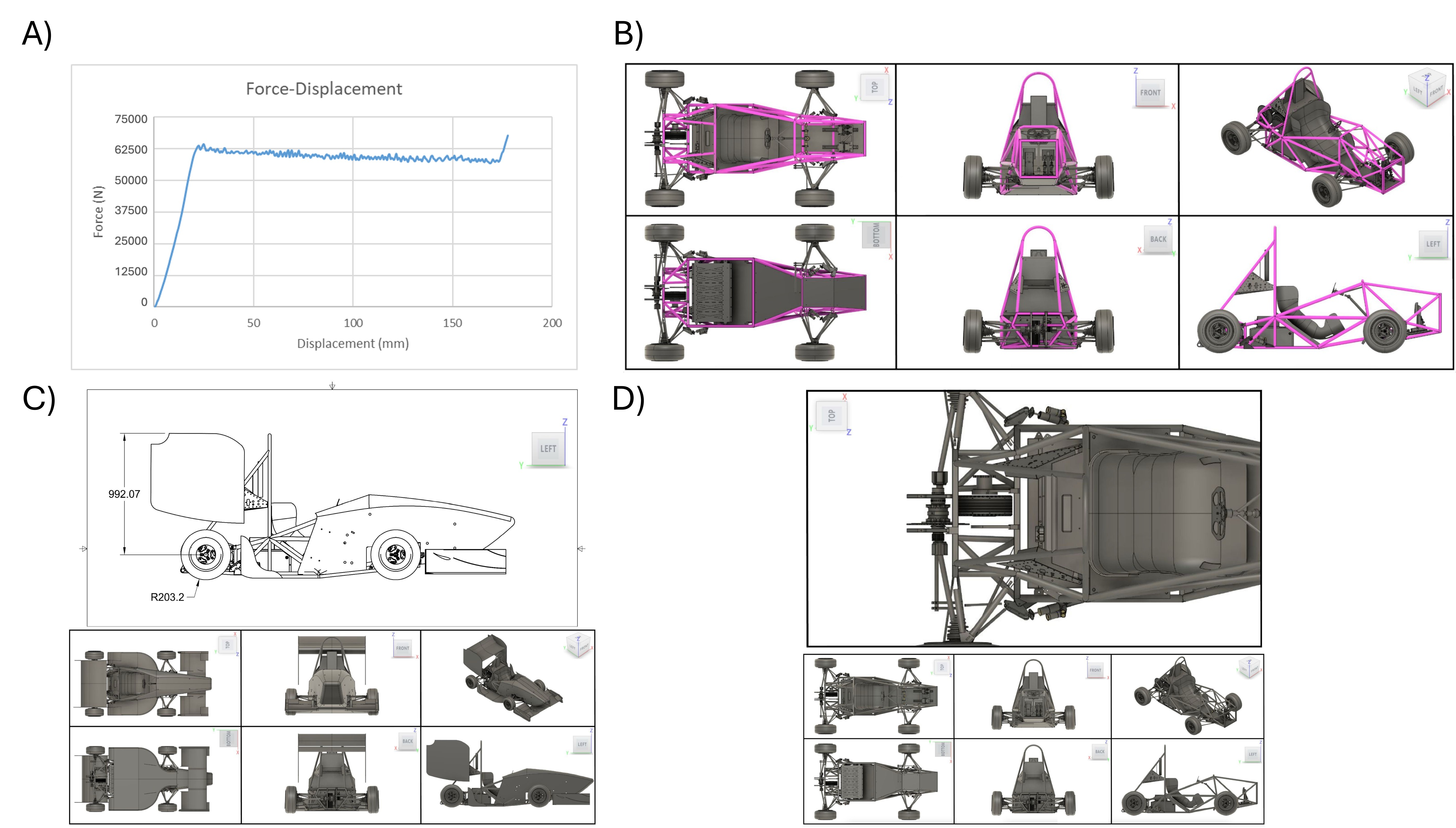}
    \caption{Visual examples of DesignQA tasks: A) Functional Performance, B) Definition, C) Dimension, and D) Presence. Images are From DesignQA Dataset~\cite{doris2025designqa}.}
    \label{fig:examplesQA}
\end{figure*}

\section{Main Prompt} \label{sec:GeneralPrompt}
The task queries come from the DesignQA benchmark dataset (sample examples are provided in Section~\ref{DE}). The main prompt used in this work (except for specialized variant prompts, which are specified within the paper) can be summarized as follows.

\textbf{System instruction.} You are an expert assistant specialising in analysing complex PDF pages and attached images. (i) Carefully read the retrieved rule pages and any extra image. (ii) Ground the answer strictly in the retrieved content.

\textbf{User message construction.} The user message consists of (i) the image inputs (if any) and (ii) the DesignQA question text appended after the image inputs.

For details on the exact model configurations and message formatting used in our implementation, please refer to the GitHub repository.

\section{Other Evaluation Metrics} \label{EM}

For the Dimension and Functional Performance subsets of DesignQA, we computed additional evaluation metrics beyond accuracy and F1 as they include an explanation part in the generated answer: BLEU-2, ROUGE-L, and Similarity scores, as discussed in DesignQA~\cite{doris2025designqa}. These metrics were intended to assess the quality of model-generated explanations compared to human-written reference explanations.

Table~\ref{tab:mllm_comparison} presents the BLEU, ROUGE, and Similarity scores for all models tested. Among these metrics, the Similarity scores, which use Sentence-BERT embeddings to compute cosine similarity, provide a rough quantitative estimation for semantic alignment, with values ranging from 0.58 to 0.78 across models.
\begin{table*}[t]
\caption{Detailed comparison of various MLLM models' scores on DesignQA benchmark (other metrics)}
\label{tab:mllm_comparison}
\centering
\footnotesize
\setlength{\tabcolsep}{3.5pt}
\begin{tabular}{llcc}
\toprule
\makecell[c]{\textbf{Category}} & \makecell[c]{\textbf{Model}} &
\makecell[c]{Dimension\\(BLEU / ROUGE / Sim. ↑)} &
\makecell[c]{Functional Perf.\\(BLEU / ROUGE / Sim. ↑)} \\
\midrule
\multirow{8}{*}{\makecell[l]{Base Models\\DesignQA~\cite{doris2025designqa}}}
& GPT-4o-AllRules & 0.18 / 0.34 / 0.78 & 0.23 / 0.41 / 0.75 \\
& GPT-4-AllRules & 0.12 / 0.30 / 0.73 & 0.17 / 0.34 / 0.70 \\
& GPT-4o-RAG & 0.11 / 0.26 / 0.64 & 0.18 / 0.37 / 0.74 \\
& GPT-4-RAG & 0.09 / 0.24 / 0.59 & 0.12 / 0.31 / 0.70 \\
& LLaVA-1.5-RAG & 0.10 / 0.24 / 0.58 & 0.16 / 0.32 / 0.65 \\
& Gemini-1.0-RAG & 0.18 / 0.34 / 0.64 & 0.27 / 0.44 / 0.73 \\
& Claude-Opus-RAG & 0.14 / 0.30 / 0.70 & 0.17 / 0.35 / 0.75 \\
\midrule
\multirow{6}{*}{\makecell[l]{Proposed\\Framework}}
& GPT-4o-MCERF-Main & 0.12 / 0.27 / 0.72 & 0.14 / 0.31 / 0.74 \\
& GPT-5-MCERF-Main & 0.15 / 0.32 / 0.74 & 0.10 / 0.26 / 0.70 \\
& GPT-5-MCERF-SelfConsistency & 0.15 / 0.31 / 0.74 & 0.08 / 0.22 / 0.68 \\
& GPT-5-MCERF-HighReasoning & 0.15 / 0.32 / 0.74 & 0.11 / 0.26 / 0.70 \\
& GPT-5-MCERF-Vision2Text & 0.11 / 0.27 / 0.68 & 0.12 / 0.28 / 0.72 \\
\bottomrule
\end{tabular}
\end{table*}
However, the BLEU and ROUGE scores sometimes reveal limitations when applied to technical engineering explanations. The scores show opposite trends compared to the accuracy and F1 scores. For MCERF, results show improved accuracy and F1 in most instances but reduced BLEU and ROUGE, and even similarity score in explanations. This discrepancy arises due to the fact that models generate explanations that are semantically accurate and technically detailed but are of a different structure, format, and wording from the human reference explanation~\cite{jourdan2025identifying}. A model might correctly identify a dimensional violation and provide valid reasoning, but use different sentence structure, alternative technical terminology, or include additional relevant details not present in the reference. The n-gram matching approach of BLEU and the longest common subsequence approach of ROUGE-L harshly penalize such differences in wording, even if the resulting engineering reasoning is correct and arrives at the right compliance decision.

Furthermore, models such as Gemini-1.0-RAG tend to produce explanations that are more closely matched to the length and surface form of the reference texts, leading to higher BLEU and ROUGE scores. However, these concise explanations sometimes lack technical detail or fail to identify the correct compliance decision, resulting in lower accuracy and F1 scores.

Given these limitations, we observe that F1 and accuracy scores remain optimal and most suitable as model performance metrics for these types of questions because they directly represent whether the model correctly identifies rule compliance, while Similarity scores provide useful complementary information about semantic alignment. We include these metrics for completeness but caution against their use as primary evaluation criteria.

\bibliographystyle{Format}

\bibliography{bibliography}

\end{document}